%% file: main.tex
\documentclass[letterpaper, 10 pt, conference]{ieeeconf}  

\IEEEoverridecommandlockouts                              
\usepackage{color}
\usepackage{amssymb}  
\usepackage{booktabs}
\usepackage{tabularx}
\usepackage{arydshln}
\newlength\subtabdist
\usepackage[utf8]{inputenc}
\usepackage{graphicx}
\usepackage{xcolor}
\usepackage{MnSymbol}
\usepackage[linesnumbered,ruled]{algorithm2e}
\usepackage{tikz}
\usepackage[utf8]{inputenc}
\usepackage{scalefnt}

\newtheorem{proposition}{Proposition}
\newtheorem{problem}{Problem}
\newtheorem{example}{Example}
\newtheorem{definition}{Definition}
\newtheorem{remark}{Remark}
\newtheorem{theorem}{Theorem}
\usepackage{pgfplots} 
\pgfplotsset{compat=newest} 
\pgfplotsset{plot coordinates/math parser=false} 
\newlength\figureheight 
\newlength\figurewidth 
\usepackage[strict=true,style=english]{csquotes}

\newcommand{\argmin}{\mathop{\mathrm{argmin}}} 

\newcommand{\argmax}{\mathop{\mathrm{argmax}}} 

\title{\LARGE \bf{
Average-based Robustness \\ for Continuous-Time Signal Temporal Logic
}}
\author{*Noushin Mehdipour$^{1}$, *Cristian-Ioan Vasile$^{2}$ and Calin Belta$^{1}$
\thanks{*These authors contributed equally. This work was partially supported at Boston University by the National Science Foundation under grants IIS-1723995, CPS-1446151, and CMMI-1400167.
${}^1$Noushin Mehdipour (noushinm@bu.edu), Calin Belta (cbelta@bu.edu) are with the Division of Systems Engineering at Boston University, Boston, and ${}^2$Cristian-Ioan Vasile (cvasile@mit.edu) is with the Laboratory for Information and Decision Systems at Massachusetts Institute of Technology, Cambridge, MA, USA.}%
}
\begin{document}
\maketitle
\thispagestyle{empty}
\pagestyle{empty}

\scalefont{1}
\renewcommand{\baselinestretch}{0.99}

\begin{abstract}
We propose a new robustness score for continuous-time Signal Temporal Logic (STL) specifications.
Instead of considering only the most severe point along the evolution of the signal,
we use average scores to extract more information from the signal,
emphasizing robust satisfaction of all the specifications' subformulae over their entire time interval domains.
We demonstrate the advantages of this new score in falsification and control synthesis problems
in systems with complex dynamics and multi-agent systems.
\end{abstract}

\section{INTRODUCTION}
\label{intro}
The increased adoption and deployment of cyber-physical systems in critical infrastructure in recent years have led to important questions about their correct functioning.
These devices embedded in our cars, planes, and homes are becoming increasingly complex.
Thus, automated tools are necessary to alleviate the need for manual design and proof of correct behavior.
Formal methods have provided approaches to specify temporal requirements of systems, formally verify whether systems satisfy given specifications, and automatically synthesize control policies that are guaranteed to be correct by construction~\cite{belta}.
Temporal logics such as Linear Temporal Logics (LTL) \cite{ltl}, Metric Temporal Logic (MTL) \cite{mtl}, Time Window Temporal Logic (TWTL) \cite{twtl} and Signal Temporal Logic (STL) \cite{stl} are popular specification languages due to their expressivity, similarity to natural language
, and an amenable structure to symbolic reasoning.

STL defines properties over continuous-time signals living in continuous spaces, and has been adopted for monitoring~\cite{stl}, falsification, and control problems such as path planning and multi-agent control with time constraints \cite{IROS}, \cite{dimos}, \cite{jana}.
One of the major advantages of STL is that it admits quantitative semantics~\cite{donze}, also known as robustness, which is interpreted as a measure of satisfaction or violation of a desired task or property.
Thus, problems involving STL can be set up as optimization of the robustness, and powerful optimization algorithms can be leveraged.

The traditional robustness introduced in~\cite{donze} uses $max$ and $min$ functions resulting in a non-differentiable function.
It only takes into account the most critical part of the signal, and, thus, induces: 1) a masking effect, where the satisfaction of other parts of the formulae do not contribute to the score, and 2) locality, where only the value of the signal at only one time point determines the score.
Both these properties have a negative impact when used in optimization problems. 
The masking effect hinders optimizers from obtaining gradient information to improve solutions, while locality results in solutions that are brittle to noise.
The traditional score was used as the objective function in an optimization problem and maximized using heuristic optimization algorithms such as Particle Swarm Optimization, Simulated Annealing and Rapidly Exploring Random Trees (RRTs) in different synthesis, falsification and control problems \cite{cdc}, \cite{SA}, \cite{rrt}.
Exact approaches in~\cite{raman,milp} encoded the temporal and Boolean constraints as Mixed Integer Linear Programming (MILP) problems and used off-the-shelf MILP solvers to maximize robustness.
Although MILP solved the issues of heuristic algorithms regarding guarantees on finding global optima, they were not scalable for large number of variables or complex temporal constraints, due to their NP-complete nature.
Another drawback of MILP implementation is the necessity of having both constraints and system dynamics be linear or linearizable.

Recent efforts to improve STL robustness focus on smoothing the $max$ and $min$ functions to employ gradient-based optimization techniques \cite{husam}, \cite{li}.
However, these approximations cause errors compared to the traditional robustness, and the  \textit{soundness} property is lost.
Another effort is refining the robustness function to include more information of the signal, rather than only its most satisfying or violating part.
In~\cite{akazaki},~\textit{averageSTL} robustness was defined using time average for temporal operators in continuous-time signals and used to solve a falsification problem.
This score did not tackle the problem with non-smooth $min$ and $max$ operations.
\cite{discrete} improved STL robustness for discrete signals by defining \textit{Discrete Average Space Robustness(DASR)} for \textit{Globally} and \textit{Until} operators.
The authors removed the non-smoothness by defining a simplified version called \textit{Discrete Simplified Average Space Robustness(DSASR)}.
However, a positive \textit{DASR} or \textit{DSASR} score did not correspond to satisfaction of the specification.
Therefore, similar to approximation methods, additional constraints were imposed to guarantee correctness.
Moreover, both these works used arithmetic average to define robustness, which, as we show in this paper, is not a good measure for all temporal and Boolean operators.

In \cite{ACC}, we defined \textit{Arithmetic-Geometric Mean(AGM)} robustness for discrete signals and showed the superiority of \textit{AGM} to the traditional robustness for control synthesis problems, and compared our gradient-based maximization to MILP implementations and approximation robustness.
Discretizing a system is not always a good idea, especially in systems with fast transient behaviors or unknown frequency.
Discretizing a system behavior with an incorrect frequency may result in missing an event or a violating behavior, e.g., a robot might collide with an obstacle in between the discrete-time instances even if it is collision free at those discrete-time moments.

In this paper, we extend the  STL quantitative score from \cite{ACC}
to continuous-time signals by defining an Arithmetic-Geometric Integral Mean (AGIM) robustness. We discuss the advantages of this new definition compared to the traditional robustness and previous works on average robustness.
Rather than merely evaluating the most satisfying or violating points, we evaluate each subformulae and at every time, highlighting both the degree of satisfaction and how frequently a specification is satisfied.
In contrast to previous works on average score where arithmetic mean was employed for some temporal operators, we refine robustness for all Boolean and temporal operators.
We use arithmetic- and product-based means to capture the importance of all outliers in signals based on the nature of the operators.
Moreover, in the proposed score, positive values correspond to satisfaction of the specification and negative values correspond to violation, showing the \textit{soundness} of our definition.
We use this score in falsification and multi-agent control synthesis problems for continuous-time systems, and compare the behavior and run-time complexity with traditional robustness.

 \section{PRELIMINARIES}
\label{preliminaries}
Let $f:\mathbb{R}^n \rightarrow \mathbb{R}$ be a real function. We define $[f]_ + = {\small \begin{cases} f & f > 0\\ 0 & \text{otherwise}\end{cases}}$ and $[f]_- = - [ - f]_+$, where $f =[f]_+ + [f]_-$.
\subsection{Signal Temporal Logics (STL)}
STL~\cite{stl} is a logic designed to specify temporal properties of continuous-time signals. 
A \textit{signal} $S:\mathbb{R}_{\ge 0}\rightarrow\mathbb{R}^n$ is a real-value function mapping each time $t \in \mathbb{R}_{\ge 0}$ to an $n$-dimensional vector $S(t)$. 
The STL syntax is defined as:
\begin{equation}
\label{eq:syntax}
\varphi:=\top \mid \mu \mid \lnot\varphi \mid \varphi_1\land\varphi_2 \mid \varphi_1\mathbf{U}_{[a,b]} \varphi_2,
\end{equation}
where $\top$ is the logical \textit{True},
$\mu$ is a \textit{predicate},
$\lnot$ and $\land$ are the Boolean \textit{negation} and \textit{conjunction} operators,
and $\mathbf{U}$ is the temporal \textit{until} operator.  
Other Boolean and temporal operators are defined as 
$\varphi_1\lor\varphi_2:=\lnot(\lnot\varphi_1\land\lnot\varphi_2)$ (disjunction),
$\mathbf{F}_{[a,b]}\varphi:=\top\mathbf{U}_{[a,b]}\varphi$ (\textit{eventually}),
and $\mathbf{G}_{[a,b]}\varphi:=\lnot\mathbf{F}_{[a,b]}\lnot\varphi$ (\textit{Globally}).
Due to brevity, in this paper we focus on $\mathbf{F}$ and $\mathbf{G}$ operators. Temporal operator $\mathbf{F}_{[a,b]}\varphi$ requires the
\enquote{specification $\varphi$ to become \textit{True} at some time in $[a,b]$}.
$\mathbf{G}_{[a,b]}\varphi$ requires \enquote{$\varphi$ to be \textit{True} at all times in $[a,b]$}.
$\varphi_1 \mathbf{U}_{[a,b]}\varphi_2$ is used to specify that \enquote{$\varphi_2$ must become \textit{True} at some time within $[a,b]$ and $\varphi_1$ must be always \textit{True} prior to that}.
A STL specification can have one or more predicates $\mu:=l(S)\geq 0$ connected by Boolean 
and temporal operators and 
$l: \mathbb{R}^n \to \mathbb{R}$ is a real, linear or nonlinear continuous function defined over values of elements of $S$. 
STL is equipped with qualitative semantics which shows \textit{whether} a signal $S$ satisfies a given specification $\varphi$ at time $t$ ($S(t) \models \varphi$) or violates it ($S(t) \nmodels \varphi$), and quantitative semantics, also known as \textit{robustness}, which measures \textit{how much} the signal is satisfying or violating the specification. 
\begin{definition}[STL Robustness] 
The robustness $\rho(\varphi,S,t)$ for formula $\varphi$ with respect to signal $S$ at time $t$ is recursively defined as \cite{donze}: 
\begin{equation}
\label{eq:org}
\begin{aligned}
\rho ( {\top ,S,t} ) &: =\rho _{\top} ,\\
\rho ( {\mu ,S,t} ) &: =  l(S(t))),\\
\rho \left( {\neg \varphi ,S,t} \right) &: =  - \rho (\varphi ,S,t),\\
\rho \left( {\varphi_1  \wedge \varphi_2 ,S,t} \right) &:= \min \left( {\rho (\varphi_1 ,S,t),\rho (\varphi_2 ,S,t)} \right),\\
\rho \left( {\varphi_1  \vee \varphi_2 ,S,t} \right) &:= \max \left( {\rho (\varphi_1 ,S,t),\rho (\varphi_2 ,S,t)} \right),\\
\rho \left( {{\mathbf{G}_{[a,b]}}\varphi ,S,t} \right) &: = \mathop {\min }\limits_{\tau \in{[t + a,t + b]}} {\rho (\varphi ,S,\tau)},\\
\rho \left( {{\mathbf{F}_{[a,b]}}\varphi ,S,t} \right) &: = \mathop {\max }\limits_{\tau \in{[t + a,t + b]}} {\rho (\varphi ,S,\tau)},
\end{aligned}
\end{equation}
where $\rho _{\top} \in \mathbb{R} \cup \{+\infty \}$ is the maximum robustness.
\end{definition}
\begin{theorem}
Robustness $\rho$ is sound, meaning that $\rho \left( {\varphi ,S,t} \right) > 0$ implies that signal $S$ satisfies $\varphi$ at time $t$, and $\rho \left( {\varphi ,S,t} \right) < 0$ implies that $S$ violates $\varphi$ at time $t$.
\end{theorem}
We denote the robustness of specification $\varphi$ at time $0$ with respect to the signal $S$ by $\rho(\varphi,S)$. We refer to this definition as traditional robustness.

\subsection{Geometric Product Integral}
The geometric integral $\prod\limits_a^b {f{{\left( x \right)}^{dx}}} $ is the continuous analog of the discrete product operator and is defined as~\cite{Bashirov2008}:

\[\prod\limits_a^b {f{{\left( x \right)}^{dx}} = \exp \left( {\int\limits_a^b {\ln f\left( x \right)dx} } \right)} \]
\section{Problem Statement}
\label{Problem Statement}
Consider a continuous-time dynamical system as:
\begin{equation}
\label{eq:dyn}
\begin{array}{c}
\dot q(t)=f(q,u),\;
q(0)=q_0,
\end{array}
\end{equation}
where $t \in \mathbb{R}_{\ge 0}$, $q(t) \in \mathbf{Q} \subseteq\mathbb{R}^n$ is the state, $u(t) \in \mathbf{U} \subseteq\mathbb{R}^m$ is the control input at time $t$, $q_0 \in \mathbf{Q}$ is the initial state, 
and $f:\mathbb{R}^n\rightarrow \mathbb{R}^n$ is locally Lipschitz. 
We denote the resulting system trajectory for the given control input $u(t)$ as $\langle q,u\rangle$. 
For system (\ref{eq:dyn}), we consider specifications given as STL formulae over predicates in its state. 
For example, the requirement that a vehicle maintains a maximum speed of $100$ over 10 minutes can be written as $\phi=G_{[0,10]} Speed \le 100$.\vspace{1pt}
\begin{problem}\label{prob:false} [Falsification] Given system~\eqref{eq:dyn} and a STL formula $\phi_f$ over predicates in the state $q$, find a control input $u(t)$ such that the resulting trajectory $\langle q,u\rangle$ violates the specification, i.e. $\langle q,u\rangle \nmodels \phi_f$.
\end{problem}\vspace{1pt}
\begin{problem}\label{prob:syn} [Synthesis] Given system~\eqref{eq:dyn} and a STL formula $\phi_s$ over predicates in the state $q$, find a control input $u(t)$ such that the resulting trajectory $\langle q,u\rangle$ satisfies the specification, i.e., $\langle q,u\rangle \models \phi_s$.
\end{problem}
In other words, a falsification problem is interpreted as finding a counterexample for the given specification to predict possible faults that may occur in system, e.g.,  falsification of $\phi$ happens if \enquote{at some time between 0 and 10, the speed goes beyond the $100$ limit}. However, in a control synthesis problem, we are interested in finding a control input such that the system trajectory meets the desired requirements, e.g., we want \enquote{the vehicle speed to be less than $100$ for all times between 0 and 10}. 
Previous works use traditional and average-based robustness to find violating (falsification) and satisfying (synthesis) trajectories for the given specification. However, as described in Sec.~\ref{intro}, traditional robustness considers only the most satisfying or violating subformula and time, and discard information of the other parts. Other average-based scores refine robustness only for the temporal operators using arithmetic means, which also have some limitations. We address these shortcomings by designing a new robustness for continuous-time signals.\vspace{1pt}

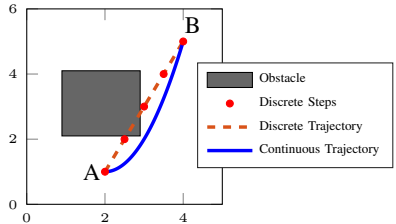
\begin{figure}[!htb]
    \centering
    \input{colision.tex} \vspace{-6pt}
    \caption{Failure in collision avoidance in a discrete-time system}
    \label{fig:col}
\end{figure}

{\bf Motivating Example:} Assume we have an agent with the specification \enquote{\textit{eventually} reach point $B$ from point $A$ and \textit{always} avoid obstacle (colored in black)}. The circle marks in Fig. \ref{fig:col} show the discrete steps the agent takes to reach $B$. Although these steps do not collide with obstacle and result in a positive discrete robustness, the trajectory connecting these steps passes through the obstacle. However, using a continuous-time score, we can correctly find a trajectory that does not collide with obstacle at any time. This example illustrates the need for a continuous-time score, as discretizing the system is not always preferable, especially when an appropriate discretization frequency is not known.
\section{ARITHMETIC-GEOMETRIC INTEGRAL MEAN (AGIM) ROBUSTNESS}

We extend our work in \cite{ACC} and propose a new average-based robustness $\eta$ for bounded continuous-time signals that captures more information about the signal relative to the traditional score. Our robustness definition returns a normalized score $\eta \in [-1,1]$ with $\eta \in (0,1]$ and $\eta \in [-1,0)$ corresponding to satisfaction and violation of the specification, respectively; and $\eta =0$ when satisfaction is inconclusive. Similar to traditional robustness, $| \eta |$ is a measure of how much the specification is satisfied or violated, while the normalization helps to have a meaningful comparison between signals of different scales.

Throughout the definitions and proofs, we assume that we have bounded signals, all \textit{Lebesgue integrable} in additive and multiplicative sense~\cite{Bashirov2008}, and the components normalized to the interval $[-1, 1]$.
\begin{definition}[AGIM Robustness] \label{def:AGIM}
Let $S: \mathbb{R}_{\ge 0} \to {[-1,1]}^n$ with $s_i$ being its $i^{th}$ component and $\pi \in [-1,1]$
. The normalized AGIM robustness $\eta (\varphi,S,t)$ 
with respect to signal $S$ at time $t$ is recursively defined as:
\begin{itemize}
\item {\bf logical $True$} \quad $\eta (\top,S ,t) : = 1$
\item {\bf $\varphi: s_i \ge \pi $} \quad 
$\eta (\varphi,S ,t) : =\frac{1}{2}( s_i(t)  - {\pi })$
\item {\bf Negation} \quad $\eta (\lnot \varphi ,S ,t) : =  - \eta(\varphi ,S ,t)$

\item {\bf Boolean and temporal operators} \quad See ~\eqref{eq:def_total}
\end{itemize}
\end{definition}\vspace{1.5pt}
Algorithm \ref{alg:rob} describes the steps to determine satisfaction or violation of specification $\phi$, and to recursively calculate the AGIM robustness with respect to signal $S$.
\begin{algorithm}[!bth]
\caption{\textsc{STL Satisfaction and AGIM Robustness Recursive calculation}}
\label{alg:rob}
\KwIn{STL Formula $\phi$; Signal $S$}
\KwOut{AGIM Robustness $\eta(\phi,S,t)$}
Find $\eta(\varphi_i,S,\tau)$ for $i=\{1,2,...,m\}$ and $\tau\in [a,b]$\; 
{\textbf{CASE}} $\phi= \varphi_1 \wedge \varphi_2 \wedge ... \varphi_m $\;
\uIf {${ANY}(\eta(\varphi_i,S,t) \le 0 )$ for $i=\{1,2,...,m\}$}{ $S  \nmodels \phi \quad$ \CommentSty{\textbf{\small{Violation}}}\;
$\eta (\phi ,S,t) := \frac{1}{m} \sum\limits_{i} {[\eta(\varphi_i,S,t)]_-}$\;}
\Else{ $S  \models \phi \quad $ \CommentSty{\textbf{\small{Satisfaction}}}\;
$\eta(\phi,S,t):=\sqrt[m]{{\prod\limits_{i = 1,...,m} {\left( {1 + \eta ({\varphi _i},S,t)} \right)} }} - 1$.}

{\textbf{CASE}} $\phi= \varphi_1 \vee \varphi_2 \vee ... \varphi_m $\;
\uIf {${ANY}(\eta(\varphi_i,S,t) > 0 )$ for $i=\{1,2,...,m\}$}{ $S \models \phi \quad$ \CommentSty{\textbf{\small{Satisfaction}}}\;
$ \eta (\phi ,S,t) := \frac{1}{m} \sum\limits_{i} {[\eta(\varphi_i,S,t)]_+}$\;}
\Else{ $S  \nmodels \phi \quad$ \CommentSty{\textbf{\small{Violation}}}\;
$\eta(\phi,S,t)=-\sqrt[m]{{\prod\limits_{i = 1,...,m} {\left( {1 - \eta ({\varphi _i},S,t)}  \right)}}} + 1$.}

{\textbf{CASE}} $\phi=\mathbf{G}_{[a,b]} \varphi$\;
\uIf {${ANY}(\eta(\varphi,S,\tau) \le 0 )$ for $\tau \in [a,b]$}{ $S  \nmodels \phi \quad$ \CommentSty{\textbf{\small{Violation}}}\;
$\eta (\phi ,S,t) := \frac{1}{b-a}{\int\limits_a^b {[\eta(\varphi, S, \tau)]_- d{\tau}}}$ \;}
\Else{ $S  \models \phi \quad$\CommentSty{\textbf{\small{Satisfaction}}}\;
$\eta(\phi,S,t):=\sqrt[b-a]{\prod\limits_a^b {{{\left( 1+ \eta ({\varphi },S,\tau) \right)}^{d\tau}}}}-1$.}

{\textbf{CASE}} $\phi=\mathbf{F}_{[a,b]}\varphi$\;
\uIf {${ANY}(\eta(\varphi_i,S,\tau) > 0 )$ for $\tau\in [a,b]$}{ $S \models \phi \quad$ \CommentSty{\textbf{\small{Satisfaction}}}\;
$ \eta (\phi ,S,t) := \frac{1}{b-a}{\int\limits_a^b {[\eta(\varphi, S, \tau)]_+ } d{\tau}}$ \;}
\Else{ $S  \nmodels \phi \quad$ \CommentSty{\textbf{\small{Violation}}}\;
$\eta(\phi,S,t)= -\sqrt[b-a]{\prod\limits_a^b {{{\left( 1- \eta ({\varphi },S,\tau) \right)}^{d\tau}}}}+1$.}
\end{algorithm}

\begin{figure*}[!htb]
\begin{equation}\label{eq:def_total}
\begin{split}
\eta ({\varphi _1} \land {\varphi _2}\land ... \land{\varphi _m} ,S ,t) := \begin{cases}
\sqrt[m]{{\prod\limits_{i = 1,...,m} {\left( {1 + \eta ({\varphi _i},S,t)} \right)} }} - 1 & \forall i \in [1,...,m]\;.\;\eta ({\varphi _i},S,t) > 0, \\
\frac{1}{m}\sum\limits_{i=1,...,m} [\eta(\varphi_i, S, t)]_-  &  \text{otherwise}
\end{cases} \\%
\eta (\varphi_1 \lor \varphi_2...\lor \varphi_m,S,t):=\begin{cases}
\frac{1}{m}\sum\limits_{i=1,...,m} [\eta(\varphi_i, S, t)]_+    & \exists i \in [1,...,m] \ .\ \eta ({\varphi}_i,S,t)>0, \\
-\sqrt[m]{{\prod\limits_{i = 1,...,m} {\left( {1 - \eta ({\varphi _i},S,t)}  \right)}}} + 1 &  \text{otherwise}
\end{cases}\\%
\eta ({{\mathbf{G}}_{[a,b]}}\varphi ,S ,t) := \begin{cases}
\sqrt[b-a]{\prod\limits_a^b {{{\left( 1+ \eta ({\varphi },S,\tau) \right)}^{d\tau}}}}-1 &  \forall \tau \in [t+a, t+b] \ .\ \eta ({\varphi},S,\tau)>0,  \\
{\frac{1}{b-a}{\int\limits_a^b {[\eta(\varphi, S, t_k')]_- } d\tau}}  & \text{otherwise}
\end{cases}\\%
\eta ({{\mathbf{F}}_{[a,b]}}\varphi ,S ,t) := \begin{cases}
{\frac{1}{b-a}{\int\limits_a^b {[\eta(\varphi, S, t_k')]_+ } d\tau}} &  \exists \tau \in [t+a, t+b] \ .\ \eta ({\varphi},S,\tau)>0,  \\
-\sqrt[b-a]{\prod\limits_a^b {{{\left( 1- \eta ({\varphi },S,\tau) \right)}^{d\tau}}}}+1  & \text{otherwise}
\end{cases}\\
\end{split}
\end{equation}
\end{figure*}

\begin{remark}
The $ANY$ method employed in Algorithm~\ref{alg:rob} is dependent on the representation
of signals and the problem solved, i.e., falsification, synthesis, verification or monitoring.
As with traditional robustness, there exist efficient representations and data structures for these tasks.
\end{remark}

\begin{theorem}[Soundness]
\label{th:soundness} The AGIM robustness is sound, i.e., a trajectory with strictly positive robustness satisfies the given specification, and a trajectory with strictly negative robustness violates the given specification:
\begin{equation}
\label{eq:sound}
\begin{aligned}
\eta(\varphi,S,t)>0 \Leftrightarrow \rho(\varphi,S,t)>0
\Rightarrow S\models\varphi,
\\
\eta(\varphi,S,t)<0 \Leftrightarrow \rho(\varphi,S,t)<0 
\Rightarrow S\not\models\varphi.\\
\end{aligned}
\end{equation}
\end{theorem}\vspace{3pt}
\begin{proof}[Sketch]
We prove the soundness by structural induction over the formula $\phi$.
The \emph{base case} corresponding to $\varphi \in \{\top, \mu\}$ is trivially true.
For the \emph{induction case}, we assume that the property holds for subformulae $\varphi$, and we need to show that it holds under Boolean and temporal operators.
For brevity, we show only the ``Globally'' case.
If $\eta(\mathbf{G}_{[a, b]}\varphi, S, t) > 0$, then $\eta(\varphi, S, \tau) > 0$ for all $\tau \in [t+a, t+b]$, which implies $\rho(\varphi, S, \tau) > 0$, $\forall\tau \in [t+a, t+b]$, and thus $\rho(\mathbf{G}_{[a, b]}\varphi, S, t) > 0$.
Conversely, if $\eta(\mathbf{G}_{[a, b]}\varphi, S, t) < 0$, then
there exists $\tau \in [t+a, t+b]$ such that $\eta(\varphi, S, \tau) < 0$, which implies $\rho(\varphi, S, \tau) < 0$ and $\rho(\mathbf{G}_{[a, b]}\varphi, S, t) < 0$.
\end{proof}\vspace{3pt}
\begin{proposition}
Let $S$ be a left (right) continuous signal and $\phi$ a STL formula.
If $\eta(\phi, S, t) = 1$ (maximum normalized score), then $\eta(\varphi, S, \tau) = 1$ for all
subformulae $\varphi$ of $\phi$ and
appropriate times $\tau$ as given by~\eqref{eq:def_total}.
Similarly,  
if $\eta(\phi, S, t) = -1$ (minimum normalized score), then $\eta(\varphi, S, \tau) = -1$ for all
subformulae $\varphi$ of $\phi$ and
appropriate times $\tau$ in~\eqref{eq:def_total}. \\Proof follows directly from Definition \ref{def:AGIM}.
\end{proposition}\vspace{1pt}

\subsection{Averaging Properties}
The AGIM robustness finds satisfaction or violation of specification $\phi$ regarding all the subformulae $\varphi$ of $\phi$ and at all appropriate times in the interval while the $min$ and $max$ functions in the traditional robustness result in a score which only considers the most critical time or subformula.
In contrast to~\cite{akazaki,discrete} where only arithmetic mean was used, we argue for the need of both arithmetic and geometric integral means for different cases as follows. The arithmetic mean is affected by the total sum value of data and is usually used when no significant outliers are present. On the other hand, the geometric mean is sensitive to unevenness and is able to measure consistency in data. Consider the \textit{eventually}, $F_{[a,b]}\varphi$, which is satisfied if $\varphi$ is satisfied at least at one time. Taking the arithmetic mean, we have a score that takes into account the total sum of all satisfying times and is also sensitive to the critical ones (outliers). Therefore, if at some time we have a large score for subformula $\varphi$, the score for $F_{[a,b]}\varphi$ is highly affected by that. On the other hand, for the \textit{globally}, $G_{[a,b]}\varphi$, to be satisfied, we need $\varphi$ to be satisfied at all times. Even a single time near violation (with small positive score) has a significant impact on the satisfaction of the specification. Therefore, for this case we will use the geometric mean to not only regard all the satisfying times, but also emphasize the consistency in satisfaction.
In other words, for $G_{[a,b]}\varphi$ to have a high score, we need all the times to have (even) high scores. The same argument holds for the robustness of the $\land$ and $\lor$ operators.
\subsection{Logical Properties}
\begin{theorem}[Boolean Property] The following hold:
\begin{enumerate}
\item {\em Idempotence:} 
$\eta(\varphi\land\varphi,S)=\eta(\varphi,S)$
\item {\em Commutativity:} 
$\eta(\varphi_1\land\varphi_2,S)=\eta(\varphi_2\land\varphi_1,S)$
\item {\em Monotonicity:} 
$\eta(\varphi_1\land\varphi_2,S)\leq\eta(\varphi_3\land\varphi_4,S),$
$\forall \varphi_i \; \text{where} \; \eta(\varphi_1,S)\leq \eta(\varphi_3,S), \; \eta(\varphi_2,S)\leq \eta(\varphi_4,S)$
\end{enumerate}
The same properties hold for disjunction $\lor$.
\end{theorem}
\vspace{3pt}
\begin{theorem}[Rules of Inference]
The following hold:
\begin{enumerate}
\item {\em Law of non-contradiction:} \vspace{-6pt}
\[\eta(\varphi \land \lnot\varphi,S) < 0 \;,\; \forall \varphi \;\;\; \text{where} \;\;\; \eta(\varphi,S)\neq 0\]
\item {\em Law of excluded middle:} \vspace{-6pt}
\[\eta(\varphi \lor \lnot\varphi,S) > 0 \;,\; \forall \varphi \;\;\; \text{where} \;\;\; \eta(\varphi,S)\neq 0\]
\item {\em Double negation:} \vspace{-6pt}
\[\eta(\lnot(\lnot\varphi),S)=\eta(\varphi,S)\;,\; \forall \varphi\]
\item {\em DeMorgan's law:}
\begin{equation*}
    \begin{array}{l}
     \eta(\varphi_1\lor\varphi_2,S)=\eta(\lnot(\lnot\varphi_1\land\lnot\varphi_2),S)\\
      \eta(\mathbf{G}_{[a,b]}\varphi,S)=\eta(\lnot\mathbf{F}_{[a,b]}\lnot\varphi,S)
    \end{array}
\end{equation*}

\end{enumerate}
All proofs follow directly from Definition \ref{def:AGIM}.
\end{theorem}

\subsection{Smoothness Properties}
The AGIM robustness $\eta(\phi,S,t)$ is smooth in $S \in [-1, 1]^n$ almost everywhere
except on the satisfaction boundaries $\rho(\varphi,S,\tau)=0$,
where $\varphi$ is a subformula of $\phi$, and appropriate times $\tau$
as given in~\eqref{eq:def_total}.
Moreover, the gradient of $\eta$ with respect to the elements of $S$ that are
part of $\phi$'s predicates is non-zero wherever it is smooth.
The AGIM robustness $\eta(\phi,S,t)$ is left-continuous in $t$ for continuous signals, and differentiable in $t$ almost everywhere if $S$ is differentiable.

\begin{proof}[Sketch]
All properties follow by structural induction.
For continuity (differentiability) in $t$, we also need to show that the set of times where the function is not continuous (differentiable) is countable.
This follows from the left-continuity, which implies only a countable set of times of discontinuity exists.
\end{proof}

\section{Robustness Optimization}
\label{optimization}
We formulate the falsification and control synthesis problems defined in  Sec.~\ref{Problem Statement} as optimization problems. Based on soundness of AGIM robustness, to find a violating trajectory for a specification $\phi_f$, we can check if $\eta(\phi_f,\langle q,u\rangle)<0$. Smaller $\eta$ corresponds to a more violating behavior. Therefore, 
we can solve the falsification Problem~\ref{prob:false} by minimizing the robustness of satisfaction of the specification $\phi_f$ over all allowed control inputs:
\begin{equation}
\label{eq: opt_fals}
\begin{array}{c}
u^*={\argmin}_u \eta(\phi_f,\langle q,u\rangle),\\
\text{s.t.}\;\;\;\; \eta(\phi_f,\langle q,u\rangle)<0,\\
\;\;\;\dot q(t)=f(q,u),\\
q(0)=q_0,\\
q(t) \in \mathbf{Q} \subseteq\mathbb{R}^n,\\
u(t) \in \mathbf{U} \subseteq\mathbb{R}^m.
\end{array}
\end{equation}
Similarly, soundness of AGIM robustness allows us to determine satisfaction of specification $\phi_s$ if $\eta(\phi_s,\langle q,u\rangle)>0$.
Larger $\eta$ corresponds to a stronger satisfaction of the desired requirements. Therefore, we can solve the synthesis Problem~\ref{prob:syn} and find the trajectory which best satisfies the desired specification $\phi_s$ by maximizing robustness over all allowed control inputs:
\begin{equation}
\label{eq: opt_syn}
\begin{array}{c}
u^*={\argmax}_u \eta(\phi_s,\langle q,u\rangle),\\
\text{s.t.}\;\;\;\; \eta(\phi_s,\langle q,u\rangle)>0,\\
\;\;\;\dot q(t)=f(q,u),\\
q(0)=q_0,\\
q(t) \in \mathbf{Q} \subseteq\mathbb{R}^n,\\
u(t) \in \mathbf{U} \subseteq\mathbb{R}^m.
\end{array}
\end{equation}
 As discussed earlier, our robustness definition is smooth and differentiable almost everywhere. In \cite{ACC}, we assumed system dynamics is also smooth and used gradient ascent to optimize the robustness. In this work, we use the MATLAB Optimization Toolbox to deal with more complex and not necessarily differentiable dynamics as in \cite{auto}. 
We focus on finding piecewise constant inputs. For a given horizon $T$, we consider the continuous-time input to be in the form of:
\begin{equation}
\label{eq:sample}
  u(t) = {u_k},\;\;\;\;\;\; (k-1){T_s} \le t \le k{T_s}  
\end{equation}
where $T_s$ is input sample time, $k \in \mathbb{N}, k \le \frac{T}{Ts}$. We hold each sample value $u_k$ constant for one sample interval $T_s$ to create a continuous-time input $u(t)$. 
We apply this continuous-time input to the system~\eqref{eq:dyn} to generate the continuous-time trajectory. 
The optimization processes for falsification and control synthesis start with generating a random sample sequence $u_s=\{u_1,u_2,...,u_{T/T_s}\}$, converted to a continuous-time input $u(t)$ using~\eqref{eq:sample} and finding system execution $\langle q,u\rangle$ starting from initial state $q_0$. We then use Matlab Constrained Parallel Optimization Toolbox to find an optimal control policy $u^*_s$ under imposed constraints which optimizes the robustness $\eta$ for the given STL constraints $\phi$. All algorithms and simulations are implemented in Matlab running on an iMac with 3.3GHz Intel Core i5 CPU 32GB RAM. 
\section{Case Studies}
\label{case}
In this section, we demonstrate the efficacy and scalability of the proposed robustness to solve the falsification and control synthesis problems defined above. We compare our results with the traditional robustness in performance and computation time.

\begin{figure}[t]
\centering
\begin{tabular}{cc}
\includegraphics[width=0.23\textwidth]{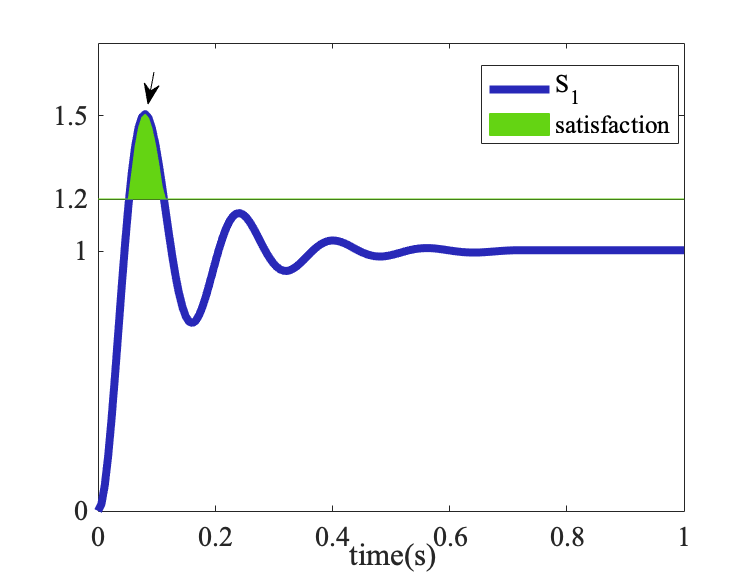} &
        \includegraphics[width=0.23\textwidth]{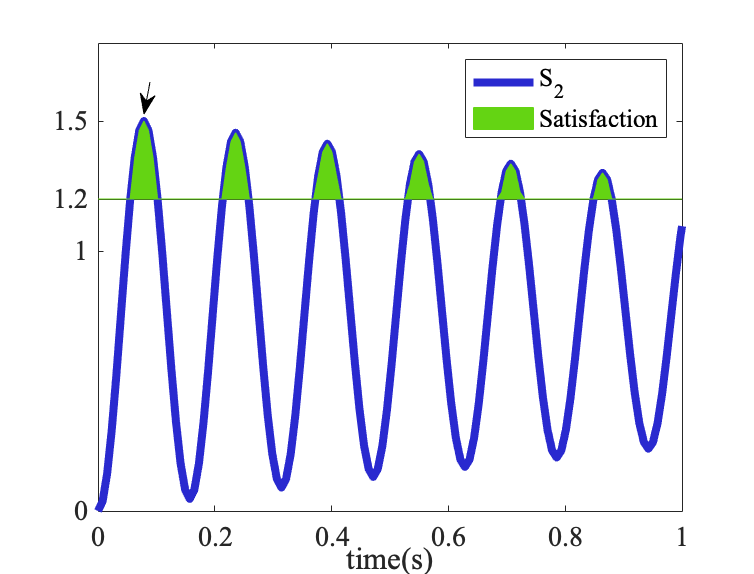}
 \end{tabular}
 \caption{Transient behavior of two dynamical systems within the first second with same $\rho$ (determined by the points marked with arrow) and different $\eta$ (determined by the areas colored in green).}
\label{fig:monitor}\vspace{-1pt} 
\end{figure}

We start with a simple verification problem, in which we compare the traditional and proposed robustness for a trajectory produced by the system under a given control input. Assume we want to study the step response of a dynamical system and consider the case where we want to find if the system response takes values greater than a threshold, say $1.2$. We can specify this behavior using STL as $\phi= F_{[0,T]} S > 1.2 $, where $S$ is the step response and $T$ is the duration time. Fig.~\ref{fig:monitor} shows the step responses of two different systems during the first second. The traditional robustness considers only the most satisfying part of the response, therefore, returns the same robustness for both systems determined by the point marked with arrow: $\rho(\phi,S_{1})=\rho(\phi,S_{2})= \mathop {{\rm{max}}}\limits_{t \in [0,T]} \left( {S_i(t) - 1.2} \right)=0.3$. However, the AGIM robustness takes the time average over the signal at all the satisfying time intervals determined by the colored area, and returns $\eta(\phi,S_{1}) \;<<\; \eta(\phi,S_{2})$ which helps to distinguish between the behaviors of the two systems. This example also illustrates the importance of having a continuous-time robustness rather than discretizing the dynamics and using a discrete-time score. For instance, if $S_1$ was discretized with a frequency smaller than $15Hz$, we would have missed the overshoot since the discrete robustness would return a non-positive score and fail to provide correct information regarding the satisfaction of the specification.
\subsection{Falsification}
\begin{figure}
    \centering
    \includegraphics[width=0.47\textwidth]{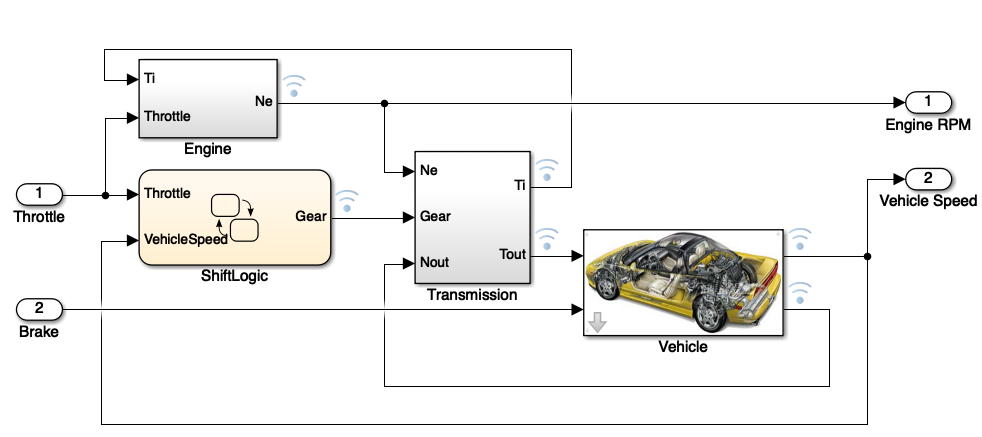}
    \caption{The Simulink automatic transmission model diagram \cite{auto}.}
    \label{fig:auto}
\end{figure}

We use the Automatic Transmission Model from Simulink \cite{auto} shown in Fig.~\ref{fig:auto}, and compare falsifying the traditional robustness versus the proposed one both in computation time and performance. We show that, by using the new robustness, we can find not only a violating execution, but a more severe violating execution which indeed requires a higher priority to be managed. This is helpful especially in the design stage to figure out the worst performance of the system for a given temporal and space constraints and limits on inputs.

\begin{figure}[t]
    \centering
\begin{tabular}{cc}
    \includegraphics[width=0.23\textwidth]{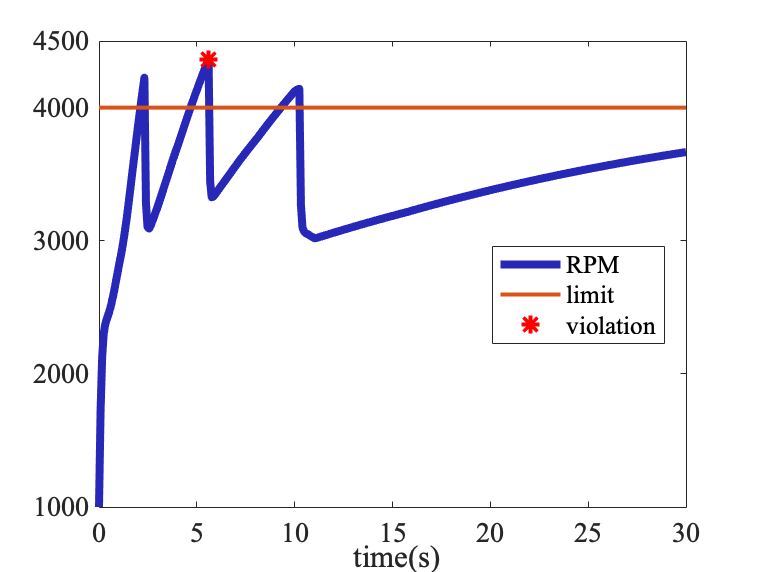} &
        \includegraphics[width=0.23\textwidth]{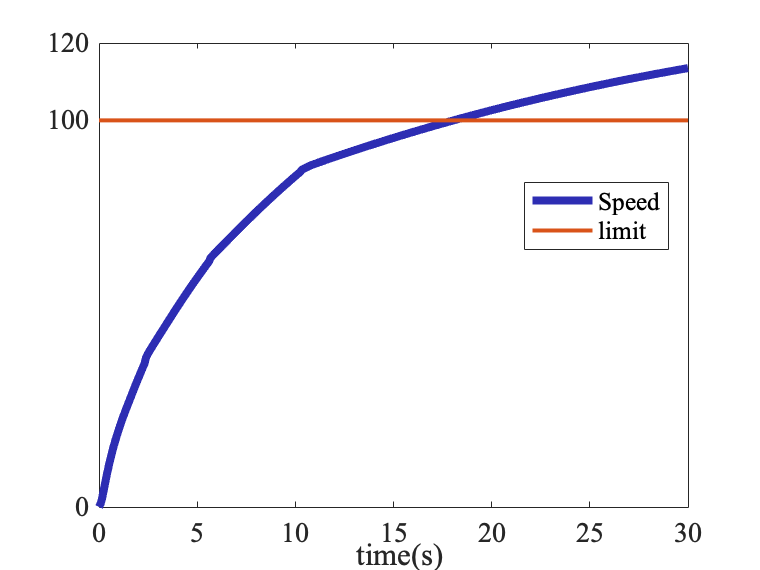}
    \end{tabular}
    \caption{Falsifying execution ($RPM$ Left, $Speed$ Right) minimizing traditional robustness $\rho$ determined by the single point marked with *.}
    \label{fig:fals_t}
\end{figure}

\begin{figure}[t]
    \centering
\begin{tabular}{cc}
    \includegraphics[width=0.23\textwidth]{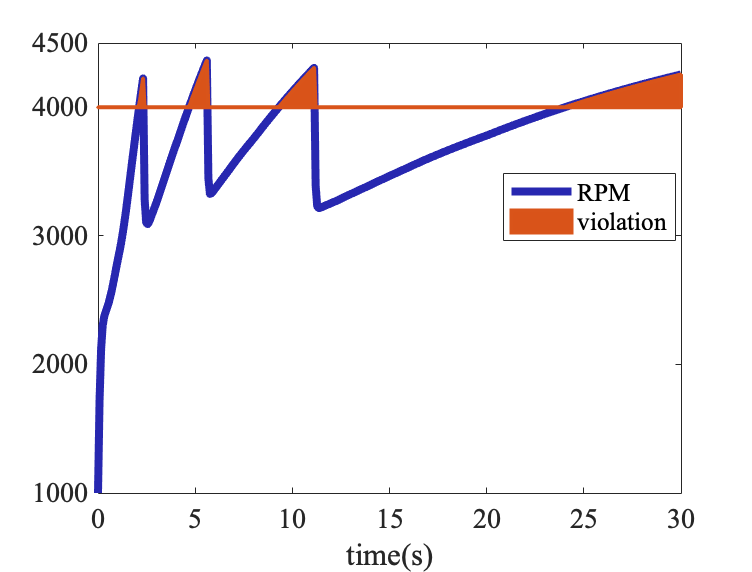} &
        \includegraphics[width=0.23\textwidth]{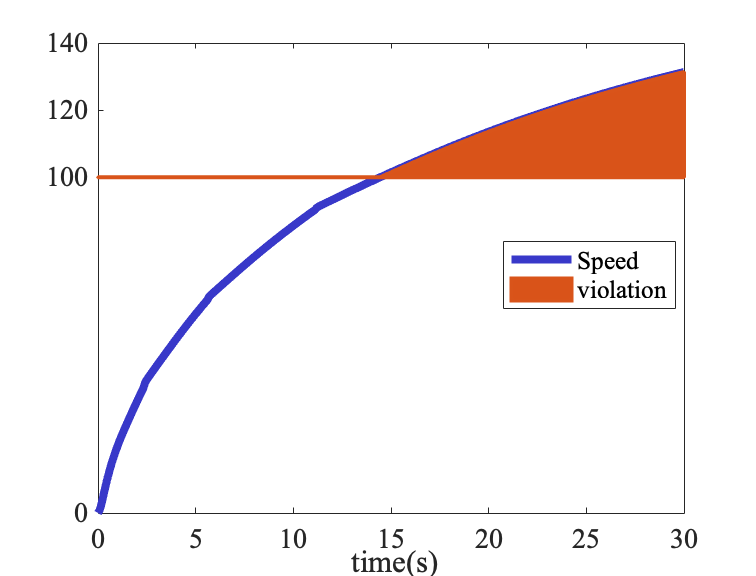}
    \end{tabular}
    \caption{Falsifying execution ($RPM$ Left, $Speed$ Right) minimizing robustness $\eta$ determined by the areas colored in red.}
    \label{fig:fals_min}
\end{figure}

In Fig.~\ref{fig:auto}, the simulation time is $T=30$ seconds and \textit{Throttle} is the input $u(t)$ with $\mathbf{U}=[0,80]$ and parameterized as a piecewise constant signal with $T_s=5$ as: 
\[u(t) = \left\{ {\begin{array}{*{20}{c}}
{{u_1}}&{0 \le t < 5}\\
{{u_2}}&{5 \le t < 10}\\
 \vdots & \vdots \\
{{u_6}}&{25 \le t < 30}
\end{array}} \right.\]
The desired requirement of the system is: \enquote{$RPM$ must always be less than $4000$ and $Speed$ must always be less than $100$ between time $0$ and $30$ seconds}, specified as:
\begin{equation}
   \phi_{Falsify}= G_{[0,30]} {RPM} \le 4000 \wedge G_{[0,30]}{Speed} \le 100
\end{equation}
The falsification for this specification happens if \enquote{$RPM$ is greater than $4000$ or $Speed$ greater than $100$}. Fig.~\ref{fig:fals_t} and Fig.~\ref{fig:fals_min} show $Speed$ and $RPM$ traces found by minimizing the traditional and the AGIM robustness, respectively. As illustrated in Fig.~\ref{fig:fals_t}, for the traditional robustness, the unnormalized score is calculated considering only the most violating part of the signal, $ \min \left(\mathop {\min }\limits_{t \in [0,30]} \left( {4000 - RPM(t)}\right),\mathop {\min }\limits_{t \in [0,30]} \left({100 - Speed(t)} \right)\right)$, marked with $*$. Note that, although the speed is violating the limit after $t=20$, robustness is only affected by $RPM$ (most violating subformula). On the other hand, the traces found by minimizing the AGIM robustness, calculated using~\eqref{eq:def_total}, evaluate all violating parts of both $RPM$ and $Speed$ over the entire time and results in a more severe violating behavior, shown as the  colored area in Fig. \ref{fig:fals_min}. Table \ref{table 1} shows the average run time, number of optimization iterations and total number of robustness evaluations to find the first falsifying traces (first time robustness is negative) and when traces with minimum robustness are found. The average time to evaluate the robustness at each evaluation is 0.34 ms for the traditional robustness and 0.41 ms for the AGIM robustness. Note that the time in the first column includes the time of generating the input, running the simulink for the generated input at each evaluation, generating resulting trajectories and calculating robustness.
\begin{table}[t]
\caption{Comparison between traditional and AGIM robustness for falsifying $\phi_{Falsify}$}
\centering
\resizebox{\columnwidth}{!}{\label{table 1}
\begin{tabular}{c c c c c c c c c}
\hline
& \multicolumn{3}{c}{\textbf{Traditional}} & 
  \multicolumn{3}{c}{\textbf{AGIM}} \\
  \addlinespace
& Time &\#Itr  & \#FuncEval  & Time & \#Itr  & \#FuncEval  \\
\hline
  \addlinespace
First Falsifying Trace & 23Sec.& 4 & 35  & 27Sec. & 7 & 56 \\
Min. Falsifying Trace & 49Sec. &14 & 107 & 54Sec. & 20 & 138\\
\hline
\end{tabular}}
\end{table}
\subsection{Control Synthesis}
We use the proposed robustness in a multi-agent system with time constraints. The agents' high level task is to achieve consensus or formation, and for certain time intervals, we impose additional temporal tasks for each agent.\vspace{1pt}
\begin{example} We consider $2$ agents with double integrator dynamics. We would like the agents to achieve consensus and meanwhile satisfy some temporal requirements. The agents' dynamics are given as:
\begin{equation}
\begin{array}{l}
{{\dot p}_i}(t) = {v_i}(t), \;\;\;\;
{{\dot v}_i}(t) = u_{{c_i}}(t)+ {u_i}(t), \;\;\;\; i=1,2
\end{array}
\end{equation}
where $p_i$ is position, $v_i$ is velocity, $u_{c_i}$ is the input to reach consensus and $u_i$ is the input to be synthesized for agent $i$ to satisfy the temporal task. The consensus input is defined as \cite{noushin}:
\begin{equation}
 {u_{{c_i}}}= { - {\gamma _p}\sum\limits_{j \in {N_i}} {{a_{ij}}({p_i} - {p_j})}  - {\gamma _v}\sum\limits_{j \in {N_i}} {{a_{ij}}({v_i} - {v_j})}  - \gamma_d {v_i}}
\end{equation}
where $N_i$ is the set of neighboring agents for $i$, $a_{ij}$ shows whether agent $i$ is connected to agent $j$, $\gamma_p,\gamma _v,\gamma_d$ are constant coefficients for consensus on position, speed and dampening the speed. The desired task is \enquote{\textit{Eventually} $Agent1$ visits \textit{Blue} $and$ $Agent2$ visits \textit{Green} within $[5,15]$ $and$ \textit{eventually} $Agent1$ and $Agent2$ visit $\textit{Yellow}$ within $[15,20]$ $and$ \textit{Always} within $[0,20]$ $Agent1$ and $Agent2$ stay inside the boundary with speeds being in the allowed range}, specified as STL formula:
\begin{equation}
\begin{array}{c}

   \phi_1= F_{[5,15]} \;p_1 \in {Blue} \;\;\wedge \;\; F_{[5,15]} \;p_2 \in {Green} \;\;\wedge \\ F_{[15,20]} \;p_1 \in {Yellow} \;\;\wedge \;\; F_{[15,20]} \;p_2 \in {Yellow} \;\; \wedge \\ G_{[0,20]} \;p_1,p_2 \in {\mathbf{P}} \;\;\wedge \;\;  G_{[0,20]} \;v_1,v_2 \in {\mathbf{V}},
\end{array}
\end{equation}
where 
$p_i=[x_i,y_i]$ is the position vector with $\mathbf{P}=[0,10]^2$ and initial states $p_{1_0}=[0,4]$, $p_{2_0}=[5,2]$, $v_i=[{vx}_i,{vy}_i]$ is the velocity with $\mathbf{V}=[-2,2]^2$ and $u_i=[u_{x_i},u_{y_i}]$ is the input vector with $\mathbf{U}=[-2,2]^2$. The regions are represented as logical formulae, for instance, for $Agent2$ visiting the $Green$ region, we have $p_2 \in Green :=6 \le x_2 \; \land \;x_2 \le 8 \;\land \;5 \le y_2 \;\land \;y_2 \le 7$.
\\
The trajectory obtained by applying the optimal control input $u^*$ to each agent found by maximizing the robustness $\eta$ with $T_s=0.1$ is shown in Fig. \ref{fig:cons} (Left). Within $[0,5]$, there is no individual temporal task for the agents except for staying inside the boundary. Therefore, the consensus input drives the agents to move towards each other. Starting at time $t=5$, each agent is supposed to eventually visit a region within the next 10 seconds. The synthesized input $u^*$ pushes the agents to visit their assigned regions \textit{as fast as possible} and \textit{stay} in each region (\textit{center}) \textit{as long as possible}, as it results in a higher score due to the averaging properties of $\eta$ over time~\eqref{eq:def_total}, and the definition of space robustness, e.g.,  $\mathop{\argmax}\limits_{x_2,y_2}\;\eta\left(p_2 \in Green,[x_2,y_2]\right)=[7,6]$ ~\eqref{eq:def_total}. Later, within $[15,20]$, both agents visit region $\textit{Yellow}$ (\textit{center}) \textit{as fast as possible} and \textit{stay} there until $t=20$. \\
We next add an obstacle to the environment, and update the specification such that both agents avoid the obstacle:
\begin{equation}
\phi_2= \phi_1 \;\wedge \; G_{[0,20]} \;p_1 \notin {Black} \; \wedge\; G_{[0,20]}\; p_2 \notin {Black}
\end{equation}
Fig. \ref{fig:cons} (Right) shows the agents' trajectories satisfying $\phi_2$ and avoiding the obstacle. Note that the trajectories are updated to avoid the obstacle, and due to the constraints on time and control input, the agents visit region $Yellow$ (robustness is positive) but do not reach its center. Fig. \ref{fig:ro_cons} shows the scores corresponding to each agent visiting the assigned regions for the specified time interval. 
In Fig.\ref{fig:ro_cons} (Left), $Agent1$ reaches $Blue$ at $t=7.35$ and stays until $t=12.9$, and reaches $Yellow$ at $t=16.35$. $Agent2$ reaches $Green$ at $t=8.9$ and stays until $t=14.2$, and reaches $Yellow$ at $t=16.4$. In Fig.\ref{fig:ro_cons} (Right), where the obstacle is added, the agents change their trajectories to avoid the obstacle. Therefore, it takes a longer time to get to $Yellow$, $Agent1$ arrives at $t=18.20$ and $Agent2$ at $t=17.10$. Note that there is no temporal task in the first $5$ seconds, and we illustrate trajectories up to $t=20$ to show the satisfaction of the temporal tasks but consensus is achieved at later times.
\end{example}
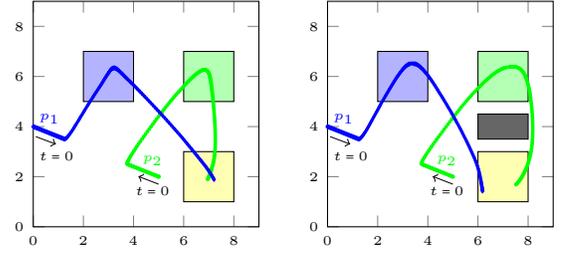
\begin{figure}[h]
    \centering
    \begin{tabular}{cc}
   \input{con.tex}  &  
   \input{obs.tex}  
    \end{tabular} \vspace{-5pt}
  \caption{Agents' trajectories satisfying $\phi_1$ (Left) and $\phi_2$ (Right)}
    \label{fig:cons}
\end{figure}
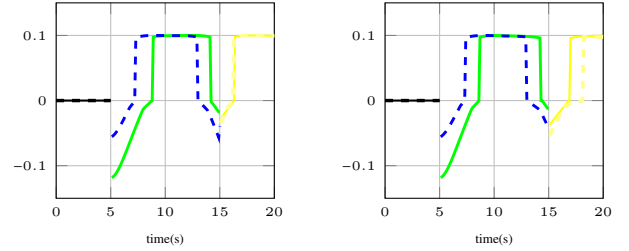
\begin{figure}[h]
    \centering
    \begin{tabular}{cc}
   \input{ro_con.tex}  &  
   \input{ro_con_obs.tex}  
    \end{tabular} \vspace{-6pt}
    \caption{Scores related to each agent visiting assigned regions for the specified time interval satisfying $\phi_1$ (Left) and $\phi_2$ (Right). Dashed and solid lines correspond to $Agent1$ and $Agent2$, respectively. Scores for each region are colored with the same color with positive score meaning that the agent is inside the region.}
    \label{fig:ro_cons}
\end{figure}
\begin{example} We consider a multi-agent system made of $3$ agents with single integrator dynamics. We would like the agents to form a triangle formation of length $2$ and meanwhile satisfy some temporal logic requirements. The agents' dynamics are given as:
\begin{equation}
\label{single}
\begin{array}{l}
{\dot p_i(t)} = {u_{{f_i}}(t)} + {u_i}(t), \;\;\;\;\; i=1,2,3\\
{u_{{f_i}}(t)}= { - \gamma_p \sum\limits_{j \in {N_i}} {{a_{ij}}({p_i}(t) - {p_j}(t)-d_{ij})} },
\end{array}
\end{equation}
where $u_i$ is the input to agent $i$ to be synthesized in order to satisfy the temporal logic requirements,  $u_{f_i}$ is the input to achieve the formation, and $d_{ij}$ is the distance between agents $i$ and $j$ \cite{olfati}.
The desired task is \enquote{\textit{Eventually} $Agent1$ visits \textit{Blue} within $[5,15]$ $and$ \textit{Eventually} $Agent2$ visits \textit{Green} within $[15,25]$ $and$ \textit{eventually} $Agent3$ visits \textit{Red} within $[25,35]$ $and$ \textit{eventually} $Agent1$ visits \textit{Yellow} within $[35,40]$ and \textit{Always} stay in \textit{Yellow} for the next $5$ seconds $and$ \textit{Always} all agents stay inside the boundary}, specified as STL formula:
\begin{equation}
\begin{array}{c}
   \phi_3= F_{[5,15]} \;p_1 \in {Blue} \;\;\wedge \;\; F_{[15,25]} \;p_2 \in {Green} \;\;\wedge \\ F_{[25,35]} \;p_3 \in {Red}\;\; \wedge F_{[35,40]}G_{[0,5]} \;p_1 \in {Yellow}\;\;\wedge \\ G_{[0,45]} \;p_1,p_2,p_3 \in {\mathbf{P}} 
\end{array}
\end{equation}
where 
$p_i=[x_i,y_i]$ is the position vector with $\mathbf{P}=[0,10]^3$ and initial states $p_{1_0}=[4,0]$, $p_{2_0}=[2,2]$, $p_{3_0}=[1,0]$ and $u_i=[u_{x_i},u_{y_i}]$ is the input vector with $\mathbf{U}=[-3,3]^3$.
\\
The trajectory obtained by applying the optimal control input $u^*$ to each agent found by maximizing the robustness $\eta$  with $T_s=0.1$ is shown in Fig. \ref{fig:form}. Within $[0,5]$, the formation input drives the agents to form a triangular formation. Starting at time $t=5$, $Agent1$ eventually visits its assigned region within the next 10 seconds (enters $Blue$ at $t=8.05$, see Fig. \ref{fig:ro_form}). Note that at this time, no temporal tasks are specified for the other agents. Therefore, only the formation input drives these agents to form a triangle. The same argument holds for $Agent2$ within $[15,25]$ and $Agent3$ within $[25,35]$. Within $[35,40]$, $Agent1$ has to visit ${Yellow}$, and stay there for at least $5$ seconds. The synthesized input $u^*$ pushes the agents to visit their assigned regions \textit{as fast as possible} and \textit{stay} in each region (\textit{center}) \textit{as long as possible} as it results in a higher score~\eqref{eq:def_total}. 
\begin{figure}[t]
    \centering
    \input{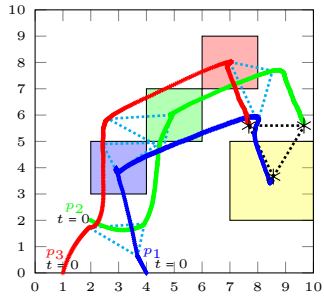}
    \caption{Agents' trajectories satisfying $\phi_3$ with agents staying in the center of assigned regions as long as possible. Triangles in cyan show the formation at $t=5$, $t=13$, $t=19$ and $t=28$ formed due to $u_{f_i}$ while the agents meet their temporal requirements, and the black triangle shows the final formation at $t=45$.}
    \label{fig:form}
\end{figure}
Fig. \ref{fig:ro_form} shows the scores related to each agent visiting the assigned regions for the specified time interval. During the first $5$ seconds, only formation input is applied (no temporal tasks assigned). At $t=8.05$, $Agent1$ enters $Blue$ and stays in its center until $t=15$, meanwhile, the formation input drives the other agents to move together to form a triangle. $Agent2$ is in the $Green$ region at $t=15$ and stays there until $t=23.1$, $Agent3$ enters $Red$ at $t=26.35$ and stays there until $t=35$, and $Agent1$ visits $Yellow$ at $t=35$ and stays there, and the desired formation is achieved. 
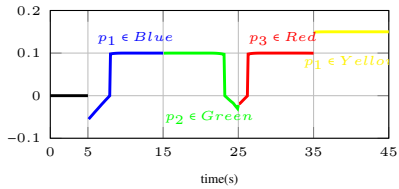
\begin{figure}[t]
    \centering
    \input{roo.tex} \vspace{-5pt} 
    \caption{Scores related to each agent visiting assigned regions for the specified time interval satisfying $\phi_3$. Scores for each region are colored with its same color, and positive score means the agent is inside the region.}
    \label{fig:ro_form}
\end{figure}
\end{example}

\section{CONCLUSION AND FUTURE WORK}
\label{sec:conclusion}
We presented a novel robustness score for continuous-time STL, which uses arithmetic and geometric integral means. We demonstrated that this score incorporates the requirements of all the subformulae and all the times of the formula. This comes in contrast with traditional approaches that consider only critical ones. We showed that our definition provides a better violation or satisfaction score in falsification and control applications. In future work, 
we will investigate how to combine our robustness score with Control Barrier Functions to find closed form control inputs.
\bibliographystyle{IEEEtran}
\bibliography{thebibliography}
\end{document}

%% file: colision.tex
%
%
\definecolor{mycolor1}{rgb}{0.00000,0.44700,0.74100}%
\definecolor{mycolor2}{rgb}{0.85000,0.32500,0.09800}%
\definecolor{mycolor3}{rgb}{0.49400,0.18400,0.55600}%
\pgfplotsset{tick label style={font=\tiny}}
\begin{tikzpicture}

\begin{axis}[%
width=2.6cm,
height=2.6cm,
at={(0cm,0cm)},
scale only axis,
xmin=0,
xmax=5,
ymin=0,
ymax=6,
axis background/.style={fill=white},
legend style={at={(.871,0.185)}, anchor=south west, legend cell align=left, align=left, draw=white!15!black}
]

\addplot[area legend, draw=black, fill=black, fill opacity=0.6]
table[row sep=crcr] {%
x	y\\
0.9	2.1\\
2.9	2.1\\
2.9	4.1\\
0.9	4.1\\
}--cycle;
\addlegendentry{\tiny{Obstacle}}

\addplot[only marks, mark=*, mark options={}, mark size=1.3000pt, draw=red, fill=red] table[row sep=crcr]{%
x	y\\
2	1\\
2.5	2\\
3	3\\
3.5	4\\
4	5\\
};

\addlegendentry{\tiny{Discrete Steps}}

\addplot [color=mycolor2, line width=1.30pt,dashed]
  table[row sep=crcr]{%
2	1\\
2.5	2\\
3	3\\
3.5	4\\
4	5\\
};
\addlegendentry{\tiny{Discrete Trajectory}}

\addplot [color=blue, line width=1.3pt]
  table[row sep=crcr]{%
2	1\\
2.01	1.0001\\
2.02	1.0004\\
2.03	1.0009\\
2.04	1.0016\\
2.05	1.0025\\
2.06	1.0036\\
2.07	1.0049\\
2.08	1.0064\\
2.09	1.0081\\
2.1	1.01\\
2.11	1.0121\\
2.12	1.0144\\
2.13	1.0169\\
2.14	1.0196\\
2.15	1.0225\\
2.16	1.0256\\
2.17	1.0289\\
2.18	1.0324\\
2.19	1.0361\\
2.2	1.04\\
2.21	1.0441\\
2.22	1.0484\\
2.23	1.0529\\
2.24	1.0576\\
2.25	1.0625\\
2.26	1.0676\\
2.27	1.0729\\
2.28	1.0784\\
2.29	1.0841\\
2.3	1.09\\
2.31	1.0961\\
2.32	1.1024\\
2.33	1.1089\\
2.34	1.1156\\
2.35	1.1225\\
2.36	1.1296\\
2.37	1.1369\\
2.38	1.1444\\
2.39	1.1521\\
2.4	1.16\\
2.41	1.1681\\
2.42	1.1764\\
2.43	1.1849\\
2.44	1.1936\\
2.45	1.2025\\
2.46	1.2116\\
2.47	1.2209\\
2.48	1.2304\\
2.49	1.2401\\
2.5	1.25\\
2.51	1.2601\\
2.52	1.2704\\
2.53	1.2809\\
2.54	1.2916\\
2.55	1.3025\\
2.56	1.3136\\
2.57	1.3249\\
2.58	1.3364\\
2.59	1.3481\\
2.6	1.36\\
2.61	1.3721\\
2.62	1.3844\\
2.63	1.3969\\
2.64	1.4096\\
2.65	1.4225\\
2.66	1.4356\\
2.67	1.4489\\
2.68	1.4624\\
2.69	1.4761\\
2.7	1.49\\
2.71	1.5041\\
2.72	1.5184\\
2.73	1.5329\\
2.74	1.5476\\
2.75	1.5625\\
2.76	1.5776\\
2.77	1.5929\\
2.78	1.6084\\
2.79	1.6241\\
2.8	1.64\\
2.81	1.6561\\
2.82	1.6724\\
2.83	1.6889\\
2.84	1.7056\\
2.85	1.7225\\
2.86	1.7396\\
2.87	1.7569\\
2.88	1.7744\\
2.89	1.7921\\
2.9	1.81\\
2.91	1.8281\\
2.92	1.8464\\
2.93	1.8649\\
2.94	1.8836\\
2.95	1.9025\\
2.96	1.9216\\
2.97	1.9409\\
2.98	1.9604\\
2.99	1.9801\\
3	2\\
3.01	2.0201\\
3.02	2.0404\\
3.03	2.0609\\
3.04	2.0816\\
3.05	2.1025\\
3.06	2.1236\\
3.07	2.1449\\
3.08	2.1664\\
3.09	2.1881\\
3.1	2.21\\
3.11	2.2321\\
3.12	2.2544\\
3.13	2.2769\\
3.14	2.2996\\
3.15	2.3225\\
3.16	2.3456\\
3.17	2.3689\\
3.18	2.3924\\
3.19	2.4161\\
3.2	2.44\\
3.21	2.4641\\
3.22	2.4884\\
3.23	2.5129\\
3.24	2.5376\\
3.25	2.5625\\
3.26	2.5876\\
3.27	2.6129\\
3.28	2.6384\\
3.29	2.6641\\
3.3	2.69\\
3.31	2.7161\\
3.32	2.7424\\
3.33	2.7689\\
3.34	2.7956\\
3.35	2.8225\\
3.36	2.8496\\
3.37	2.8769\\
3.38	2.9044\\
3.39	2.9321\\
3.4	2.96\\
3.41	2.9881\\
3.42	3.0164\\
3.43	3.0449\\
3.44	3.0736\\
3.45	3.1025\\
3.46	3.1316\\
3.47	3.1609\\
3.48	3.1904\\
3.49	3.2201\\
3.5	3.25\\
3.51	3.2801\\
3.52	3.3104\\
3.53	3.3409\\
3.54	3.3716\\
3.55	3.4025\\
3.56	3.4336\\
3.57	3.4649\\
3.58	3.4964\\
3.59	3.5281\\
3.6	3.56\\
3.61	3.5921\\
3.62	3.6244\\
3.63	3.6569\\
3.64	3.6896\\
3.65	3.7225\\
3.66	3.7556\\
3.67	3.7889\\
3.68	3.8224\\
3.69	3.8561\\
3.7	3.89\\
3.71	3.9241\\
3.72	3.9584\\
3.73	3.9929\\
3.74	4.0276\\
3.75	4.0625\\
3.76	4.0976\\
3.77	4.1329\\
3.78	4.1684\\
3.79	4.2041\\
3.8	4.24\\
3.81	4.2761\\
3.82	4.3124\\
3.83	4.3489\\
3.84	4.3856\\
3.85	4.4225\\
3.86	4.4596\\
3.87	4.4969\\
3.88	4.5344\\
3.89	4.5721\\
3.9	4.61\\
3.91	4.6481\\
3.92	4.6864\\
3.93	4.7249\\
3.94	4.7636\\
3.95	4.8025\\
3.96	4.8416\\
3.97	4.8809\\
3.98	4.9204\\
3.99	4.9601\\
4	5\\
};
\addlegendentry{\tiny{Continuous Trajectory}}

\node[right, align=left]
at (axis cs:3.8,5.5) {\small B};

\node[right, align=left]
at (axis cs:1.2,1) { \small A};

\end{axis}

\end{tikzpicture}%

%% file: con.tex
%
%
\definecolor{mycolor1}{rgb}{0.00000,0.44700,0.74100}%
\definecolor{mycolor2}{rgb}{0.92900,0.69400,0.12500}%
\definecolor{mycolor3}{rgb}{1.00000,1.00000,0.00000}%
\pgfplotsset{tick label style={font=\tiny}}
\begin{tikzpicture}

\begin{axis}[%
width=3cm,
height=3cm,
at={(0,0)},
scale only axis,
xmin=0,
xmax=9,
ymin=0,
ymax=9,
axis background/.style={fill=white},
legend style={at={(0.147,0.129)}, anchor=south west, legend cell align=left, align=left, draw=white!15!black}
]
\addplot [color=mycolor1]
  table[row sep=crcr]{%
5	2\\
5	2\\
4.995	2.002\\
4.9857	2.00572\\
4.972712	2.0109152\\
4.95657092	2.017371632\\
4.9377441672	2.02490233312\\
4.916640017952	2.0333439928192\\
4.89361496126432	2.04255401549427\\
4.86898013247701	2.0524079470092\\
4.8430069497974	2.06279722008104\\
4.81593205242798	2.07362717902881\\
4.78796162679068	2.08481534928373\\
4.75927519663775	2.0962899213449\\
4.73002894345264	2.10798842261894\\
4.70035861532018	2.11985655387193\\
4.67038207523936	2.13184716990426\\
4.64020153353921	2.14391938658432\\
4.6099055035266	2.15603779858936\\
4.57957051464868	2.16817179414053\\
4.54926261320661	2.18029495471736\\
4.51903867693714	2.19238452922514\\
4.48894756651898	2.20442097339241\\
4.45903113420549	2.21638754631781\\
4.42932510728285	2.22826995708686\\
4.39985986186097	2.24005605525561\\
4.37066110058358	2.25173555976657\\
4.34175044616131	2.26329982153548\\
4.31314596115699	2.27474161553721\\
4.28486260316095	2.28605495873562\\
4.25691262336204	2.29723495065519\\
4.22930591552866	2.30827763378854\\
4.20205032154522	2.31917987138191\\
4.17515189888841	2.32993924044464\\
4.14861515476047	2.34055393809581\\
4.12244325101266	2.35102269959494\\
4.09663818348002	2.36134472660799\\
4.07120093889992	2.37151962444003\\
4.04613163219408	2.38154734712237\\
4.02142962654926	2.3914281493803\\
3.99709363843032	2.40116254462787\\
3.97312182939494	2.41075126824203\\
3.94951188634765	2.42019524546094\\
3.92626109166819	2.42949556333273\\
3.90336638447116	2.43865344621154\\
3.88082441409838	2.44767023436065\\
3.85863158680884	2.45654736527647\\
3.83678410651164	2.46528635739534\\
3.81527801028244	2.47388879588703\\
3.79410919931229	2.48235632027508\\
3.77327346585741	2.49069061365704\\
3.75276651668758	2.49889339332497\\
3.74617135354218	2.52075691145989\\
3.75175940162457	2.55561863153056\\
3.76780332504615	2.59993374327403\\
3.79241178290734	2.65304829858268\\
3.82507809821174	2.71472631523207\\
3.86380437521293	2.7846030899639\\
3.90928062289125	2.85942480845819\\
3.96051589975293	2.94023619261726\\
4.01532712879922	3.02678121268167\\
4.07361802396335	3.11786795138138\\
4.13585858385246	3.21122102964336\\
4.20084270419464	3.30622867379022\\
4.26875959721391	3.40364011988891\\
4.33849803067871	3.50216714187503\\
4.40933731178534	3.60110216410376\\
4.48123822782078	3.70064244197214\\
4.55427157406931	3.80061282410589\\
4.62717094746069	3.90100960181717\\
4.70028203598066	4.0020787729805\\
4.77297985905678	4.10225132597503\\
4.84596170637895	4.20103693200264\\
4.91904010821394	4.29935399366569\\
4.9919148858675	4.39580964175663\\
5.06469876058584	4.49162920058561\\
5.13766510569483	4.5863087720563\\
5.20991263823877	4.67941953318319\\
5.28070015034546	4.77005750263125\\
5.35064575986968	4.85837810843588\\
5.4201665507745	4.94409775231997\\
5.48807376877055	5.02807423945497\\
5.55622309614059	5.10925482778669\\
5.62315923221603	5.18752174373475\\
5.68809347950529	5.26223179010574\\
5.7522291764066	5.33411557471548\\
5.81541566742618	5.40422841875603\\
5.87790009108007	5.47219950098686\\
5.93787725420145	5.5380937966958\\
5.99565458748009	5.60158280854586\\
6.05127212635818	5.66234536320767\\
6.10489762777794	5.71972088415033\\
6.1567822448414	5.77362209977525\\
6.20667816165029	5.82501602192863\\
6.25472153635267	5.87309100670284\\
6.30097779980803	5.91834564092777\\
6.345827031905	5.96087200983967\\
6.3890336441647	6.00075873568817\\
6.43078467995624	6.03841774753984\\
6.47101844893424	6.07232010261427\\
6.50983469697475	6.10374075183106\\
6.54648728152166	6.13194125827757\\
6.58194069152774	6.15720383757214\\
6.61487619383528	6.17981488745371\\
6.64531742979277	6.19992206756511\\
6.67359026289905	6.21663719755255\\
6.69992004184352	6.23079822717982\\
6.72579121801521	6.24336126276152\\
6.7508212729308	6.25261084861481\\
6.77491254772724	6.25926276747883\\
6.79818447784797	6.2639115010322\\
6.82035581702383	6.26627857972579\\
6.84155740408849	6.26524015039707\\
6.86243625402816	6.26242206644225\\
6.88203299459652	6.25651339951569\\
6.90123949892886	6.24880652970189\\
6.91978460438667	6.2377777094315\\
6.93737784351495	6.22436208634682\\
6.9539392511224	6.20779349973095\\
6.96866158875526	6.18874574782061\\
6.98254894210151	6.16781317263582\\
6.99482749628043	6.14394063037727\\
7.00634498461074	6.11786376736124\\
7.01753310792592	6.08834024289489\\
7.02843782745433	6.05671006700686\\
7.03857964341472	6.02244866027685\\
7.04773642903985	5.98523670798093\\
7.05693317894545	5.94518186478782\\
7.06521846922803	5.9024230464689\\
7.07267447199368	5.85754616639136\\
7.0797092955385	5.81008698639368\\
7.0874111009525	5.76011949686539\\
7.09581519151676	5.70692537120299\\
7.10342923615987	5.65077896639859\\
7.11086241380631	5.59154039921843\\
7.11846528212204	5.52937069442875\\
7.12670420934033	5.46483145537758\\
7.13488791919385	5.39648901213298\\
7.1437069056951	5.32581128107698\\
7.15218682969009	5.25308157436254\\
7.16090279083214	5.17779752903587\\
7.16922529896352	5.09929810083029\\
7.17680931754495	5.01901747611982\\
7.18363223858484	4.93573394983542\\
7.19015332716166	4.85070697404292\\
7.19731129756658	4.7632609376066\\
7.2052292923206	4.67343573641427\\
7.21306545938829	4.58200522842766\\
7.21986004560084	4.4892744683726\\
7.22663481932738	4.39564141449866\\
7.23230900528901	4.30113656925144\\
7.23823670581736	4.20660199168185\\
7.24361441827213	4.11223955233253\\
7.24870304817841	4.01716658612944\\
7.25277475028672	3.9224608077165\\
7.25622632815088	3.82896583657805\\
7.25816098451797	3.73615759901334\\
7.26029003136029	3.64361229577456\\
7.26146168395734	3.55231895339982\\
7.26152106769154	3.46186790964432\\
7.26090890988934	3.37261728865717\\
7.25924934940697	3.28454973308028\\
7.25728804967814	3.19900462703017\\
7.25476617048809	3.11519973617936\\
7.24965078276175	3.032595377834\\
7.24226674255277	2.95131560458556\\
7.23461722029271	2.87217541768866\\
7.22500900506825	2.79560415109994\\
7.21390752449903	2.72243525093203\\
7.20105721748265	2.65128901383284\\
7.18669638589991	2.58268535491074\\
7.17189079460335	2.51599946976504\\
7.15557834368655	2.45182167126315\\
7.13872966976463	2.38977904621609\\
7.12145686096099	2.33144907129725\\
7.10485281929174	2.27573837843564\\
7.08841282732068	2.22312667614457\\
7.07185502001194	2.17342417992098\\
7.05547259500005	2.12698109998083\\
7.03881657828758	2.08302185080185\\
7.02189992469144	2.04095524253857\\
7.00595401792273	2.0019625854199\\
6.9899839428636	1.96455150313981\\
6.97376746736312	1.92965560678806\\
6.95832385782526	1.89702066833383\\
};

\addplot[only marks, mark=*, mark options={}, mark size=0.52361pt, color=green, fill=green] table[row sep=crcr]{%
x	y\\
5	2\\
5	2\\
4.995	2.002\\
4.9857	2.00572\\
4.972712	2.0109152\\
4.95657092	2.017371632\\
4.9377441672	2.02490233312\\
4.916640017952	2.0333439928192\\
4.89361496126432	2.04255401549427\\
4.86898013247701	2.0524079470092\\
4.8430069497974	2.06279722008104\\
4.81593205242798	2.07362717902881\\
4.78796162679068	2.08481534928373\\
4.75927519663775	2.0962899213449\\
4.73002894345264	2.10798842261894\\
4.70035861532018	2.11985655387193\\
4.67038207523936	2.13184716990426\\
4.64020153353921	2.14391938658432\\
4.6099055035266	2.15603779858936\\
4.57957051464868	2.16817179414053\\
4.54926261320661	2.18029495471736\\
4.51903867693714	2.19238452922514\\
4.48894756651898	2.20442097339241\\
4.45903113420549	2.21638754631781\\
4.42932510728285	2.22826995708686\\
4.39985986186097	2.24005605525561\\
4.37066110058358	2.25173555976657\\
4.34175044616131	2.26329982153548\\
4.31314596115699	2.27474161553721\\
4.28486260316095	2.28605495873562\\
4.25691262336204	2.29723495065519\\
4.22930591552866	2.30827763378854\\
4.20205032154522	2.31917987138191\\
4.17515189888841	2.32993924044464\\
4.14861515476047	2.34055393809581\\
4.12244325101266	2.35102269959494\\
4.09663818348002	2.36134472660799\\
4.07120093889992	2.37151962444003\\
4.04613163219408	2.38154734712237\\
4.02142962654926	2.3914281493803\\
3.99709363843032	2.40116254462787\\
3.97312182939494	2.41075126824203\\
3.94951188634765	2.42019524546094\\
3.92626109166819	2.42949556333273\\
3.90336638447116	2.43865344621154\\
3.88082441409838	2.44767023436065\\
3.85863158680884	2.45654736527647\\
3.83678410651164	2.46528635739534\\
3.81527801028244	2.47388879588703\\
3.79410919931229	2.48235632027508\\
3.77327346585741	2.49069061365704\\
3.75276651668758	2.49889339332497\\
3.74617135354218	2.52075691145989\\
3.75175940162457	2.55561863153056\\
3.76780332504615	2.59993374327403\\
3.79241178290734	2.65304829858268\\
3.82507809821174	2.71472631523207\\
3.86380437521293	2.7846030899639\\
3.90928062289125	2.85942480845819\\
3.96051589975293	2.94023619261726\\
4.01532712879922	3.02678121268167\\
4.07361802396335	3.11786795138138\\
4.13585858385246	3.21122102964336\\
4.20084270419464	3.30622867379022\\
4.26875959721391	3.40364011988891\\
4.33849803067871	3.50216714187503\\
4.40933731178534	3.60110216410376\\
4.48123822782078	3.70064244197214\\
4.55427157406931	3.80061282410589\\
4.62717094746069	3.90100960181717\\
4.70028203598066	4.0020787729805\\
4.77297985905678	4.10225132597503\\
4.84596170637895	4.20103693200264\\
4.91904010821394	4.29935399366569\\
4.9919148858675	4.39580964175663\\
5.06469876058584	4.49162920058561\\
5.13766510569483	4.5863087720563\\
5.20991263823877	4.67941953318319\\
5.28070015034546	4.77005750263125\\
5.35064575986968	4.85837810843588\\
5.4201665507745	4.94409775231997\\
5.48807376877055	5.02807423945497\\
5.55622309614059	5.10925482778669\\
5.62315923221603	5.18752174373475\\
5.68809347950529	5.26223179010574\\
5.7522291764066	5.33411557471548\\
5.81541566742618	5.40422841875603\\
5.87790009108007	5.47219950098686\\
5.93787725420145	5.5380937966958\\
5.99565458748009	5.60158280854586\\
6.05127212635818	5.66234536320767\\
6.10489762777794	5.71972088415033\\
6.1567822448414	5.77362209977525\\
6.20667816165029	5.82501602192863\\
6.25472153635267	5.87309100670284\\
6.30097779980803	5.91834564092777\\
6.345827031905	5.96087200983967\\
6.3890336441647	6.00075873568817\\
6.43078467995624	6.03841774753984\\
6.47101844893424	6.07232010261427\\
6.50983469697475	6.10374075183106\\
6.54648728152166	6.13194125827757\\
6.58194069152774	6.15720383757214\\
6.61487619383528	6.17981488745371\\
6.64531742979277	6.19992206756511\\
6.67359026289905	6.21663719755255\\
6.69992004184352	6.23079822717982\\
6.72579121801521	6.24336126276152\\
6.7508212729308	6.25261084861481\\
6.77491254772724	6.25926276747883\\
6.79818447784797	6.2639115010322\\
6.82035581702383	6.26627857972579\\
6.84155740408849	6.26524015039707\\
6.86243625402816	6.26242206644225\\
6.88203299459652	6.25651339951569\\
6.90123949892886	6.24880652970189\\
6.91978460438667	6.2377777094315\\
6.93737784351495	6.22436208634682\\
6.9539392511224	6.20779349973095\\
6.96866158875526	6.18874574782061\\
6.98254894210151	6.16781317263582\\
6.99482749628043	6.14394063037727\\
7.00634498461074	6.11786376736124\\
7.01753310792592	6.08834024289489\\
7.02843782745433	6.05671006700686\\
7.03857964341472	6.02244866027685\\
7.04773642903985	5.98523670798093\\
7.05693317894545	5.94518186478782\\
7.06521846922803	5.9024230464689\\
7.07267447199368	5.85754616639136\\
7.0797092955385	5.81008698639368\\
7.0874111009525	5.76011949686539\\
7.09581519151676	5.70692537120299\\
7.10342923615987	5.65077896639859\\
7.11086241380631	5.59154039921843\\
7.11846528212204	5.52937069442875\\
7.12670420934033	5.46483145537758\\
7.13488791919385	5.39648901213298\\
7.1437069056951	5.32581128107698\\
7.15218682969009	5.25308157436254\\
7.16090279083214	5.17779752903587\\
7.16922529896352	5.09929810083029\\
7.17680931754495	5.01901747611982\\
7.18363223858484	4.93573394983542\\
7.19015332716166	4.85070697404292\\
7.19731129756658	4.7632609376066\\
7.2052292923206	4.67343573641427\\
7.21306545938829	4.58200522842766\\
7.21986004560084	4.4892744683726\\
7.22663481932738	4.39564141449866\\
7.23230900528901	4.30113656925144\\
7.23823670581736	4.20660199168185\\
7.24361441827213	4.11223955233253\\
7.24870304817841	4.01716658612944\\
7.25277475028672	3.9224608077165\\
7.25622632815088	3.82896583657805\\
7.25816098451797	3.73615759901334\\
7.26029003136029	3.64361229577456\\
7.26146168395734	3.55231895339982\\
7.26152106769154	3.46186790964432\\
7.26090890988934	3.37261728865717\\
7.25924934940697	3.28454973308028\\
7.25728804967814	3.19900462703017\\
7.25476617048809	3.11519973617936\\
7.24965078276175	3.032595377834\\
7.24226674255277	2.95131560458556\\
7.23461722029271	2.87217541768866\\
7.22500900506825	2.79560415109994\\
7.21390752449903	2.72243525093203\\
7.20105721748265	2.65128901383284\\
7.18669638589991	2.58268535491074\\
7.17189079460335	2.51599946976504\\
7.15557834368655	2.45182167126315\\
7.13872966976463	2.38977904621609\\
7.12145686096099	2.33144907129725\\
7.10485281929174	2.27573837843564\\
7.08841282732068	2.22312667614457\\
7.07185502001194	2.17342417992098\\
7.05547259500005	2.12698109998083\\
7.03881657828758	2.08302185080185\\
7.02189992469144	2.04095524253857\\
7.00595401792273	2.0019625854199\\
6.9899839428636	1.96455150313981\\
6.97376746736312	1.92965560678806\\
6.95832385782526	1.89702066833383\\
};

\addplot [color=mycolor2]
  table[row sep=crcr]{%
0	4\\
0	4\\
0.005	3.998\\
0.0143	3.99428\\
0.027288	3.9890848\\
0.04342908	3.982628368\\
0.0622558328	3.97509766688\\
0.083359982048	3.9666560071808\\
0.10638503873568	3.95744598450573\\
0.131019867522989	3.9475920529908\\
0.156993050202603	3.93720277991896\\
0.184067947572025	3.92637282097119\\
0.212038373209323	3.91518465071627\\
0.240724803362255	3.9037100786551\\
0.269971056547358	3.89201157738106\\
0.299641384679822	3.88014344612807\\
0.329617924760647	3.86815283009574\\
0.359798466460796	3.85608061341568\\
0.390094496473403	3.84396220141064\\
0.420429485351324	3.83182820585947\\
0.450737386793389	3.81970504528264\\
0.480961323062862	3.80761547077486\\
0.511052433481022	3.79557902660759\\
0.540968865794514	3.78361245368219\\
0.570674892717155	3.77173004291314\\
0.600140138139037	3.75994394474439\\
0.629338899416422	3.74826444023343\\
0.658249553838694	3.73670017846452\\
0.686854038843016	3.72525838446279\\
0.715137396839055	3.71394504126438\\
0.743087376637963	3.70276504934481\\
0.770694084471345	3.69172236621146\\
0.797949678454778	3.68082012861809\\
0.824848101111588	3.67006075955536\\
0.851384845239535	3.65944606190419\\
0.877556748987346	3.64897730040506\\
0.903361816519984	3.63865527339201\\
0.928799061100078	3.62848037555997\\
0.95386836780592	3.61845265287763\\
0.978570373450743	3.6085718506197\\
1.00290636156968	3.59883745537213\\
1.02687817060506	3.58924873175797\\
1.05048811365235	3.57980475453906\\
1.07373890833181	3.57050443666727\\
1.09663361552884	3.56134655378846\\
1.11917558590163	3.55232976563935\\
1.14136841319116	3.54345263472353\\
1.16321589348836	3.53471364260466\\
1.18472198971757	3.52611120411297\\
1.20589080068771	3.51764367972492\\
1.22672653414259	3.50930938634296\\
1.24723348331242	3.50110660667503\\
1.27251085410935	3.50720203948645\\
1.30216128592514	3.5257335883244\\
1.33545145587209	3.55577226401841\\
1.37192046533211	3.59361988399182\\
1.41114006098183	3.64108814419042\\
1.45225978925276	3.69648066492191\\
1.49561535305488	3.75640954952028\\
1.54103375213303	3.82278870071996\\
1.5878790483741	3.89369083185659\\
1.63559821271892	3.9694765265914\\
1.68410061308682	4.04754502943041\\
1.73320109105936	4.12676733193081\\
1.78298638884105	4.20805349650862\\
1.83305531902179	4.28987741633113\\
1.88332303677067	4.3733434436805\\
1.93340832374592	4.45652861794105\\
1.98427135255165	4.54088409878652\\
2.03534805606398	4.62527578515832\\
2.08653718053072	4.70863831702285\\
2.13805923039688	4.79194128239844\\
2.18893792901089	4.87525085037775\\
2.23920271883557	4.95666145682773\\
2.28970545061662	5.03619973720014\\
2.33956186911451	5.11318763891052\\
2.38856451639959	5.18801590207132\\
2.43651666647698	5.26197885880337\\
2.48393998346081	5.33288751265951\\
2.52868201454586	5.40240715806289\\
2.57302535661018	5.46906841481178\\
2.61437447961402	5.53393406292736\\
2.65527420943736	5.5951625483665\\
2.69358249222866	5.65490356828136\\
2.72968358092833	5.71228693260295\\
2.76376059385462	5.76656882040378\\
2.79587642997066	5.8188257091732\\
2.82616222967633	5.8691779254344\\
2.85541552013675	5.91666150432899\\
2.88279868033253	5.96091507841709\\
2.90940660378296	6.00256347058652\\
2.93467097698898	6.04135294480337\\
2.95835968548972	6.07699038428215\\
2.98015611590262	6.10983681419565\\
3.00069595171273	6.14072371046569\\
3.01964886258886	6.1698928804031\\
3.03801171975389	6.19678957615286\\
3.05613946691491	6.22042843649702\\
3.07306092984353	6.24222217016601\\
3.08962554091455	6.26197536683994\\
3.10582471329004	6.27948813647548\\
3.12222776692146	6.29511778730665\\
3.1388881787441	6.30799354635174\\
3.15507228237584	6.31868145711522\\
3.17177689197474	6.32752554856788\\
3.18886763556946	6.33426779373048\\
3.20564667478791	6.33913450956966\\
3.22303095783533	6.34041218412582\\
3.24044686124514	6.33901242926393\\
3.25788880583565	6.33485718406073\\
3.27644850198651	6.32739712827973\\
3.29563375898963	6.31835839514071\\
3.31641234665028	6.30682559172271\\
3.33766018802955	6.29358736384592\\
3.35955906176977	6.27809327175302\\
3.38298820422307	6.25974536108334\\
3.40764590202992	6.23880427640984\\
3.43324397394903	6.21491792493419\\
3.4607805513777	6.18943883423154\\
3.48969082022002	6.16159608392898\\
3.5202747501403	6.1320615452875\\
3.5524256655737	6.10038819605881\\
3.58620433121808	6.06667466056989\\
3.6214577805919	6.02999214172765\\
3.66046391043934	5.99055401538062\\
3.70257771082433	5.94880841085074\\
3.74782217946867	5.90513107867413\\
3.79598409507785	5.85918068368456\\
3.84670504039059	5.8095361593698\\
3.89990992329319	5.75776757963254\\
3.95557926088145	5.70322569663281\\
4.0133213036491	5.64559966535623\\
4.07283361767426	5.58626122587476\\
4.13385608163356	5.5234324617799\\
4.19745523326144	5.45746264177356\\
4.26344125812708	5.38964207836462\\
4.33068619564992	5.3190662955728\\
4.40001135878843	5.24610477055742\\
4.47162612348297	5.17098733873307\\
4.5445030846155	5.09440243674244\\
4.61909115619179	5.01533390800952\\
4.6954260677668	4.9348384330307\\
4.7728785643424	4.8531587374724\\
4.85087709830672	4.76958984701752\\
4.9299236112465	4.68490860783028\\
5.0095152588972	4.59927155532521\\
5.08943478753784	4.51269152039353\\
5.17041698868509	4.42463399184317\\
5.25140529717494	4.33503618691994\\
5.33310216835011	4.24356464719688\\
5.41590316854507	4.15035539733925\\
5.49842280185362	4.05760867230243\\
5.58023006559438	3.9655291453257\\
5.66223417928298	3.87329895849762\\
5.74360942102295	3.78135306034376\\
5.82411447550984	3.69078673831875\\
5.9017384073139	3.60086090102536\\
5.97656929559821	3.51143541920468\\
6.0497860616638	3.42270302235844\\
6.12191754052241	3.33638014155625\\
6.19170076859681	3.25249083057556\\
6.25968283909639	3.16964187926163\\
6.32540610064854	3.08932548935648\\
6.38855801206705	3.01186060090594\\
6.44946078978335	2.93694148511884\\
6.50890024945572	2.86338926100984\\
6.56693676247844	2.79272292113882\\
6.62212084964051	2.72438660660779\\
6.67576993372706	2.65883010347645\\
6.7269836183681	2.59511548512527\\
6.77623011436025	2.53380909056052\\
6.82234314273665	2.47440509975522\\
6.86676888525336	2.41698915173901\\
6.9081190686738	2.36172171365255\\
6.94692339427328	2.30845737156612\\
6.98353524809776	2.25702952344694\\
7.01746750652572	2.20716901190999\\
7.04903852091606	2.15948992126322\\
7.07784180982375	2.11406325693066\\
7.10514488703428	2.07028630252466\\
7.12995908929419	2.02903803159677\\
7.15257462639917	1.9900596074523\\
7.17343610882045	1.9529734067293\\
7.19347975810015	1.91737521554005\\
7.2112303629081	1.88377887362424\\
};

\addplot[only marks, mark=diamond*, mark options={}, mark size=0.6584pt, color=blue, fill=blue] table[row sep=crcr]{%
x	y\\
0	4\\
0	4\\
0.005	3.998\\
0.0143	3.99428\\
0.027288	3.9890848\\
0.04342908	3.982628368\\
0.0622558328	3.97509766688\\
0.083359982048	3.9666560071808\\
0.10638503873568	3.95744598450573\\
0.131019867522989	3.9475920529908\\
0.156993050202603	3.93720277991896\\
0.184067947572025	3.92637282097119\\
0.212038373209323	3.91518465071627\\
0.240724803362255	3.9037100786551\\
0.269971056547358	3.89201157738106\\
0.299641384679822	3.88014344612807\\
0.329617924760647	3.86815283009574\\
0.359798466460796	3.85608061341568\\
0.390094496473403	3.84396220141064\\
0.420429485351324	3.83182820585947\\
0.450737386793389	3.81970504528264\\
0.480961323062862	3.80761547077486\\
0.511052433481022	3.79557902660759\\
0.540968865794514	3.78361245368219\\
0.570674892717155	3.77173004291314\\
0.600140138139037	3.75994394474439\\
0.629338899416422	3.74826444023343\\
0.658249553838694	3.73670017846452\\
0.686854038843016	3.72525838446279\\
0.715137396839055	3.71394504126438\\
0.743087376637963	3.70276504934481\\
0.770694084471345	3.69172236621146\\
0.797949678454778	3.68082012861809\\
0.824848101111588	3.67006075955536\\
0.851384845239535	3.65944606190419\\
0.877556748987346	3.64897730040506\\
0.903361816519984	3.63865527339201\\
0.928799061100078	3.62848037555997\\
0.95386836780592	3.61845265287763\\
0.978570373450743	3.6085718506197\\
1.00290636156968	3.59883745537213\\
1.02687817060506	3.58924873175797\\
1.05048811365235	3.57980475453906\\
1.07373890833181	3.57050443666727\\
1.09663361552884	3.56134655378846\\
1.11917558590163	3.55232976563935\\
1.14136841319116	3.54345263472353\\
1.16321589348836	3.53471364260466\\
1.18472198971757	3.52611120411297\\
1.20589080068771	3.51764367972492\\
1.22672653414259	3.50930938634296\\
1.24723348331242	3.50110660667503\\
1.27251085410935	3.50720203948645\\
1.30216128592514	3.5257335883244\\
1.33545145587209	3.55577226401841\\
1.37192046533211	3.59361988399182\\
1.41114006098183	3.64108814419042\\
1.45225978925276	3.69648066492191\\
1.49561535305488	3.75640954952028\\
1.54103375213303	3.82278870071996\\
1.5878790483741	3.89369083185659\\
1.63559821271892	3.9694765265914\\
1.68410061308682	4.04754502943041\\
1.73320109105936	4.12676733193081\\
1.78298638884105	4.20805349650862\\
1.83305531902179	4.28987741633113\\
1.88332303677067	4.3733434436805\\
1.93340832374592	4.45652861794105\\
1.98427135255165	4.54088409878652\\
2.03534805606398	4.62527578515832\\
2.08653718053072	4.70863831702285\\
2.13805923039688	4.79194128239844\\
2.18893792901089	4.87525085037775\\
2.23920271883557	4.95666145682773\\
2.28970545061662	5.03619973720014\\
2.33956186911451	5.11318763891052\\
2.38856451639959	5.18801590207132\\
2.43651666647698	5.26197885880337\\
2.48393998346081	5.33288751265951\\
2.52868201454586	5.40240715806289\\
2.57302535661018	5.46906841481178\\
2.61437447961402	5.53393406292736\\
2.65527420943736	5.5951625483665\\
2.69358249222866	5.65490356828136\\
2.72968358092833	5.71228693260295\\
2.76376059385462	5.76656882040378\\
2.79587642997066	5.8188257091732\\
2.82616222967633	5.8691779254344\\
2.85541552013675	5.91666150432899\\
2.88279868033253	5.96091507841709\\
2.90940660378296	6.00256347058652\\
2.93467097698898	6.04135294480337\\
2.95835968548972	6.07699038428215\\
2.98015611590262	6.10983681419565\\
3.00069595171273	6.14072371046569\\
3.01964886258886	6.1698928804031\\
3.03801171975389	6.19678957615286\\
3.05613946691491	6.22042843649702\\
3.07306092984353	6.24222217016601\\
3.08962554091455	6.26197536683994\\
3.10582471329004	6.27948813647548\\
3.12222776692146	6.29511778730665\\
3.1388881787441	6.30799354635174\\
3.15507228237584	6.31868145711522\\
3.17177689197474	6.32752554856788\\
3.18886763556946	6.33426779373048\\
3.20564667478791	6.33913450956966\\
3.22303095783533	6.34041218412582\\
3.24044686124514	6.33901242926393\\
3.25788880583565	6.33485718406073\\
3.27644850198651	6.32739712827973\\
3.29563375898963	6.31835839514071\\
3.31641234665028	6.30682559172271\\
3.33766018802955	6.29358736384592\\
3.35955906176977	6.27809327175302\\
3.38298820422307	6.25974536108334\\
3.40764590202992	6.23880427640984\\
3.43324397394903	6.21491792493419\\
3.4607805513777	6.18943883423154\\
3.48969082022002	6.16159608392898\\
3.5202747501403	6.1320615452875\\
3.5524256655737	6.10038819605881\\
3.58620433121808	6.06667466056989\\
3.6214577805919	6.02999214172765\\
3.66046391043934	5.99055401538062\\
3.70257771082433	5.94880841085074\\
3.74782217946867	5.90513107867413\\
3.79598409507785	5.85918068368456\\
3.84670504039059	5.8095361593698\\
3.89990992329319	5.75776757963254\\
3.95557926088145	5.70322569663281\\
4.0133213036491	5.64559966535623\\
4.07283361767426	5.58626122587476\\
4.13385608163356	5.5234324617799\\
4.19745523326144	5.45746264177356\\
4.26344125812708	5.38964207836462\\
4.33068619564992	5.3190662955728\\
4.40001135878843	5.24610477055742\\
4.47162612348297	5.17098733873307\\
4.5445030846155	5.09440243674244\\
4.61909115619179	5.01533390800952\\
4.6954260677668	4.9348384330307\\
4.7728785643424	4.8531587374724\\
4.85087709830672	4.76958984701752\\
4.9299236112465	4.68490860783028\\
5.0095152588972	4.59927155532521\\
5.08943478753784	4.51269152039353\\
5.17041698868509	4.42463399184317\\
5.25140529717494	4.33503618691994\\
5.33310216835011	4.24356464719688\\
5.41590316854507	4.15035539733925\\
5.49842280185362	4.05760867230243\\
5.58023006559438	3.9655291453257\\
5.66223417928298	3.87329895849762\\
5.74360942102295	3.78135306034376\\
5.82411447550984	3.69078673831875\\
5.9017384073139	3.60086090102536\\
5.97656929559821	3.51143541920468\\
6.0497860616638	3.42270302235844\\
6.12191754052241	3.33638014155625\\
6.19170076859681	3.25249083057556\\
6.25968283909639	3.16964187926163\\
6.32540610064854	3.08932548935648\\
6.38855801206705	3.01186060090594\\
6.44946078978335	2.93694148511884\\
6.50890024945572	2.86338926100984\\
6.56693676247844	2.79272292113882\\
6.62212084964051	2.72438660660779\\
6.67576993372706	2.65883010347645\\
6.7269836183681	2.59511548512527\\
6.77623011436025	2.53380909056052\\
6.82234314273665	2.47440509975522\\
6.86676888525336	2.41698915173901\\
6.9081190686738	2.36172171365255\\
6.94692339427328	2.30845737156612\\
6.98353524809776	2.25702952344694\\
7.01746750652572	2.20716901190999\\
7.04903852091606	2.15948992126322\\
7.07784180982375	2.11406325693066\\
7.10514488703428	2.07028630252466\\
7.12995908929419	2.02903803159677\\
7.15257462639917	1.9900596074523\\
7.17343610882045	1.9529734067293\\
7.19347975810015	1.91737521554005\\
7.2112303629081	1.88377887362424\\
};

\addplot[area legend, draw=black, fill=blue, fill opacity=0.3]
table[row sep=crcr] {%
x	y\\
4	7\\
2	7\\
2	5\\
4	5\\
}--cycle;

\addplot[area legend, draw=black, fill=green, fill opacity=0.3]
table[row sep=crcr] {%
x	y\\
8	7\\
6	7\\
6	5\\
8	5\\
}--cycle;

\addplot[area legend, draw=black, fill=mycolor3, fill opacity=0.3]
table[row sep=crcr] {%
x	y\\
8	3\\
6	3\\
6	1\\
8	1\\
}--cycle;
\node[right, align=left]
at (axis cs:-0.11368482,2.8336007503) {\textcolor{black}{\tiny$t=0$}};
\node[right, align=left]
at (axis cs:-0.11368482,4.3336007503) {\textcolor{blue}{\tiny$p_1$}};
\node[right, align=left]
at (axis cs:3.71368482,1.436007503) {\textcolor{black}{\tiny$t=0$}};
\node[right, align=left]
at (axis cs:3.991368482,2.636007503) {\textcolor{green}{\tiny$p_2$}};
\draw [->, line width=0.1mm] (0.1,3.6) -- (.9,3.3);
\draw [->, line width=0.1mm] (5,1.7) -- (4.2,2.01);
\end{axis}
\end{tikzpicture}%

%% file: obs.tex
%
%
\definecolor{mycolor1}{rgb}{0.00000,0.44700,0.74100}%
\definecolor{mycolor2}{rgb}{0.92900,0.69400,0.12500}%
\definecolor{mycolor3}{rgb}{1.00000,1.00000,0.00000}%
\pgfplotsset{tick label style={font=\tiny}}
\begin{tikzpicture}

\begin{axis}[%
width=3cm,
height=3cm,
at={(0.764in,0.541in)},
scale only axis,
xmin=0,
xmax=9,
ymin=0,
ymax=9,
axis background/.style={fill=white},
legend style={at={(0.154,0.124)}, anchor=south west, legend cell align=left, align=left, draw=white!15!black}
]
\addplot [color=mycolor1]
  table[row sep=crcr]{%
5	2\\
5	2\\
4.995	2.002\\
4.9857	2.00572\\
4.972712	2.0109152\\
4.95657092	2.017371632\\
4.9377441672	2.02490233312\\
4.916640017952	2.0333439928192\\
4.89361496126432	2.04255401549427\\
4.86898013247701	2.0524079470092\\
4.8430069497974	2.06279722008104\\
4.81593205242798	2.07362717902881\\
4.78796162679068	2.08481534928373\\
4.75927519663775	2.0962899213449\\
4.73002894345264	2.10798842261894\\
4.70035861532018	2.11985655387193\\
4.67038207523936	2.13184716990426\\
4.64020153353921	2.14391938658432\\
4.6099055035266	2.15603779858936\\
4.57957051464868	2.16817179414053\\
4.54926261320661	2.18029495471736\\
4.51903867693714	2.19238452922514\\
4.48894756651898	2.20442097339241\\
4.45903113420549	2.21638754631781\\
4.42932510728285	2.22826995708686\\
4.39985986186097	2.24005605525561\\
4.37066110058358	2.25173555976657\\
4.34175044616131	2.26329982153548\\
4.31314596115699	2.27474161553721\\
4.28486260316095	2.28605495873562\\
4.25691262336204	2.29723495065519\\
4.22930591552866	2.30827763378854\\
4.20205032154522	2.31917987138191\\
4.17515189888841	2.32993924044464\\
4.14861515476047	2.34055393809581\\
4.12244325101266	2.35102269959494\\
4.09663818348002	2.36134472660799\\
4.07120093889992	2.37151962444003\\
4.04613163219408	2.38154734712237\\
4.02142962654926	2.3914281493803\\
3.99709363843032	2.40116254462787\\
3.97312182939494	2.41075126824203\\
3.94951188634765	2.42019524546094\\
3.92626109166819	2.42949556333273\\
3.90336638447116	2.43865344621154\\
3.88082441409838	2.44767023436065\\
3.85863158680884	2.45654736527647\\
3.83678410651164	2.46528635739534\\
3.81527801028244	2.47388879588703\\
3.79410919931229	2.48235632027508\\
3.77327346585741	2.49069061365704\\
3.75276651668758	2.49889339332497\\
3.74639747805482	2.52047068090392\\
3.75241287473315	2.55468375676429\\
3.76906905115753	2.59830958632723\\
3.79446414787219	2.65066635527511\\
3.82806272012688	2.71145026750972\\
3.86786164318822	2.78024747517795\\
3.91451698017558	2.85408365446456\\
3.96703273567496	2.93384793093368\\
4.02322954400446	3.01925238660352\\
4.08300337405101	3.10918948447805\\
4.14681212690457	3.20155026823066\\
4.21345209175019	3.29573651925104\\
4.28310710852751	3.39242154046985\\
4.35466843592644	3.49038667768583\\
4.42741309140902	3.58894527323136\\
4.50130446398406	3.68827258130337\\
4.57641812752438	3.788190898484\\
4.65147809227628	3.88868768656994\\
4.72684375588674	3.98998762721638\\
4.8018860625883	4.09058156265628\\
4.87732572328611	4.1899843867079\\
4.95298174428326	4.28908863073815\\
5.02855905375098	4.38652292523467\\
5.10418516471975	4.48349120378527\\
5.18015122927336	4.57949656505551\\
5.25554997794159	4.67411279961872\\
5.32963103421339	4.76643155907956\\
5.40305007808843	4.85660598932079\\
5.47628240184779	4.94434517719299\\
5.54811334746536	5.03051679756306\\
5.62050167907518	5.11404908204639\\
5.69196554237226	5.19481845020321\\
5.76172984404912	5.27215713715415\\
5.83109274535578	5.34682099337372\\
5.89994438101191	5.41989939687489\\
5.96858798587062	5.49101495060122\\
6.03518770572965	5.56024062900646\\
6.1001187779401	5.62724343740869\\
6.16346767490825	5.69169265401606\\
6.22544884605847	5.75289744864646\\
6.28635981429692	5.81076333074147\\
6.34596874269418	5.86630202894285\\
6.40443557695061	5.91865308932949\\
6.46183426010394	5.96832915922927\\
6.51856033995644	6.01541003894447\\
6.57435147636603	6.05996875848187\\
6.62938254328323	6.10241557342758\\
6.68355953776455	6.14111018001786\\
6.73695308507679	6.17737400786665\\
6.78873525081613	6.21039926786752\\
6.83988239833507	6.24045808872132\\
6.88895284984818	6.26782727432967\\
6.93592200124154	6.29263817935461\\
6.98108484613534	6.31392419922061\\
7.02462799207996	6.33255382072333\\
7.06806957379536	6.34952330882212\\
7.14095148591403	6.36299333223662\\
7.25311651054567	6.38371097921542\\
7.29463657382957	6.38229748055058\\
7.33516010202824	6.38844880734044\\
7.37477236164448	6.39096302848208\\
7.41410548357828	6.39157038161352\\
7.45209166150219	6.38887539932815\\
7.4896245598148	6.38426046841884\\
7.52636830555193	6.37610766789762\\
7.56196695939843	6.36542608167772\\
7.59628933109246	6.35141146699185\\
7.62843329421367	6.33480042458758\\
7.65942277929081	6.31624721830761\\
7.68839515855733	6.29464618275875\\
7.71621329680585	6.27080833916095\\
7.74330032195222	6.24343185591267\\
7.76967186573492	6.21397815480894\\
7.79475875319205	6.18191597120382\\
7.81827320846652	6.14693466386867\\
7.84126115489176	6.10918134723529\\
7.86267370389375	6.06883689149748\\
7.88256721612847	6.02656100006321\\
7.90135113619425	5.98188787660453\\
7.92016386207987	5.93492700689184\\
7.93902832353609	5.88493872985545\\
7.95634728938111	5.83224224357102\\
7.97276505860274	5.77671132680702\\
7.98865465115154	5.71853621187703\\
8.0045174958825	5.65832891808119\\
8.01962796416757	5.59457085052173\\
8.03473663191994	5.52882032257481\\
8.04882503160557	5.46137227626483\\
8.06253519965102	5.39166112706744\\
8.07522838732887	5.31895493162149\\
8.08657366480032	5.24474194330057\\
8.09658883928233	5.1676879521338\\
8.10581118519349	5.08908504141711\\
8.11529096690746	5.00817216378909\\
8.12520966184533	4.92494344212321\\
8.13471306372439	4.84017338350991\\
8.14281589450768	4.75414227902758\\
8.15064039635074	4.66723200445991\\
8.15706313531407	4.57943530073991\\
8.16354920614295	4.49160983969415\\
8.16926158082654	4.40393227273474\\
8.1744930553164	4.31542304179814\\
8.17847921776104	4.22719141330531\\
8.18164802464726	4.14009861206032\\
8.1830503215154	4.05354174349971\\
8.18453007705655	3.96702317370087\\
8.18487871539408	3.88155367990196\\
8.1839379809475	3.79664376687453\\
8.18217094186396	3.712622099953\\
8.17916731640544	3.62941058737184\\
8.17569389381948	3.54837968520431\\
8.17144442836054	3.46861959749976\\
8.16568413925905	3.38948427372301\\
8.15852296541455	3.3110086731817\\
8.15041028930217	3.23398424103764\\
8.1409820693201	3.15879997116761\\
8.13051877260217	3.08628283157761\\
8.11855471715271	3.01490058112001\\
8.10515249631196	2.94515781471642\\
8.09125184195626	2.87634347766468\\
8.07557641149851	2.80905996732272\\
8.05901030096639	2.74290733704193\\
8.04156006699496	2.6795774711473\\
8.02428528394902	2.61794114818865\\
8.0065809532266	2.55855360935992\\
7.98809409645775	2.50127066284259\\
7.96909138686323	2.44653099259768\\
7.94907099142181	2.39358683203726\\
7.9280283781611	2.34189270424254\\
7.90725233250464	2.2927858109104\\
7.88569380694937	2.24477317379716\\
7.86312676890134	2.19893357541186\\
7.84063949651005	2.15508831167047\\
7.81923471833377	2.11288317349388\\
7.79848455520224	2.07230062575276\\
7.77732746505362	2.03253868824919\\
7.75634675434418	1.99448384892862\\
7.73463169500801	1.9584549866464\\
7.71281632984263	1.92469251111738\\
7.69050393933314	1.89327469233256\\
7.6677838346924	1.86302565201948\\
7.64487697441893	1.83312804228392\\
7.62314832185006	1.80366785671299\\
7.60151368119774	1.77659703885802\\
7.57933900616676	1.75128805938269\\
7.5584340316474	1.72799635002207\\
7.53837112778426	1.70543318825948\\
7.51839681739248	1.68391892889235\\
};

\addplot[only marks, mark=*, mark options={}, mark size=0.502361pt, color=green, fill=green] table[row sep=crcr]{%
x	y\\
5	2\\
5	2\\
4.995	2.002\\
4.9857	2.00572\\
4.972712	2.0109152\\
4.95657092	2.017371632\\
4.9377441672	2.02490233312\\
4.916640017952	2.0333439928192\\
4.89361496126432	2.04255401549427\\
4.86898013247701	2.0524079470092\\
4.8430069497974	2.06279722008104\\
4.81593205242798	2.07362717902881\\
4.78796162679068	2.08481534928373\\
4.75927519663775	2.0962899213449\\
4.73002894345264	2.10798842261894\\
4.70035861532018	2.11985655387193\\
4.67038207523936	2.13184716990426\\
4.64020153353921	2.14391938658432\\
4.6099055035266	2.15603779858936\\
4.57957051464868	2.16817179414053\\
4.54926261320661	2.18029495471736\\
4.51903867693714	2.19238452922514\\
4.48894756651898	2.20442097339241\\
4.45903113420549	2.21638754631781\\
4.42932510728285	2.22826995708686\\
4.39985986186097	2.24005605525561\\
4.37066110058358	2.25173555976657\\
4.34175044616131	2.26329982153548\\
4.31314596115699	2.27474161553721\\
4.28486260316095	2.28605495873562\\
4.25691262336204	2.29723495065519\\
4.22930591552866	2.30827763378854\\
4.20205032154522	2.31917987138191\\
4.17515189888841	2.32993924044464\\
4.14861515476047	2.34055393809581\\
4.12244325101266	2.35102269959494\\
4.09663818348002	2.36134472660799\\
4.07120093889992	2.37151962444003\\
4.04613163219408	2.38154734712237\\
4.02142962654926	2.3914281493803\\
3.99709363843032	2.40116254462787\\
3.97312182939494	2.41075126824203\\
3.94951188634765	2.42019524546094\\
3.92626109166819	2.42949556333273\\
3.90336638447116	2.43865344621154\\
3.88082441409838	2.44767023436065\\
3.85863158680884	2.45654736527647\\
3.83678410651164	2.46528635739534\\
3.81527801028244	2.47388879588703\\
3.79410919931229	2.48235632027508\\
3.77327346585741	2.49069061365704\\
3.75276651668758	2.49889339332497\\
3.74639747805482	2.52047068090392\\
3.75241287473315	2.55468375676429\\
3.76906905115753	2.59830958632723\\
3.79446414787219	2.65066635527511\\
3.82806272012688	2.71145026750972\\
3.86786164318822	2.78024747517795\\
3.91451698017558	2.85408365446456\\
3.96703273567496	2.93384793093368\\
4.02322954400446	3.01925238660352\\
4.08300337405101	3.10918948447805\\
4.14681212690457	3.20155026823066\\
4.21345209175019	3.29573651925104\\
4.28310710852751	3.39242154046985\\
4.35466843592644	3.49038667768583\\
4.42741309140902	3.58894527323136\\
4.50130446398406	3.68827258130337\\
4.57641812752438	3.788190898484\\
4.65147809227628	3.88868768656994\\
4.72684375588674	3.98998762721638\\
4.8018860625883	4.09058156265628\\
4.87732572328611	4.1899843867079\\
4.95298174428326	4.28908863073815\\
5.02855905375098	4.38652292523467\\
5.10418516471975	4.48349120378527\\
5.18015122927336	4.57949656505551\\
5.25554997794159	4.67411279961872\\
5.32963103421339	4.76643155907956\\
5.40305007808843	4.85660598932079\\
5.47628240184779	4.94434517719299\\
5.54811334746536	5.03051679756306\\
5.62050167907518	5.11404908204639\\
5.69196554237226	5.19481845020321\\
5.76172984404912	5.27215713715415\\
5.83109274535578	5.34682099337372\\
5.89994438101191	5.41989939687489\\
5.96858798587062	5.49101495060122\\
6.03518770572965	5.56024062900646\\
6.1001187779401	5.62724343740869\\
6.16346767490825	5.69169265401606\\
6.22544884605847	5.75289744864646\\
6.28635981429692	5.81076333074147\\
6.34596874269418	5.86630202894285\\
6.40443557695061	5.91865308932949\\
6.46183426010394	5.96832915922927\\
6.51856033995644	6.01541003894447\\
6.57435147636603	6.05996875848187\\
6.62938254328323	6.10241557342758\\
6.68355953776455	6.14111018001786\\
6.73695308507679	6.17737400786665\\
6.78873525081613	6.21039926786752\\
6.83988239833507	6.24045808872132\\
6.88895284984818	6.26782727432967\\
6.93592200124154	6.29263817935461\\
6.98108484613534	6.31392419922061\\
7.02462799207996	6.33255382072333\\
7.06806957379536	6.34952330882212\\
7.14095148591403	6.36299333223662\\
7.25311651054567	6.38371097921542\\
7.29463657382957	6.38229748055058\\
7.33516010202824	6.38844880734044\\
7.37477236164448	6.39096302848208\\
7.41410548357828	6.39157038161352\\
7.45209166150219	6.38887539932815\\
7.4896245598148	6.38426046841884\\
7.52636830555193	6.37610766789762\\
7.56196695939843	6.36542608167772\\
7.59628933109246	6.35141146699185\\
7.62843329421367	6.33480042458758\\
7.65942277929081	6.31624721830761\\
7.68839515855733	6.29464618275875\\
7.71621329680585	6.27080833916095\\
7.74330032195222	6.24343185591267\\
7.76967186573492	6.21397815480894\\
7.79475875319205	6.18191597120382\\
7.81827320846652	6.14693466386867\\
7.84126115489176	6.10918134723529\\
7.86267370389375	6.06883689149748\\
7.88256721612847	6.02656100006321\\
7.90135113619425	5.98188787660453\\
7.92016386207987	5.93492700689184\\
7.93902832353609	5.88493872985545\\
7.95634728938111	5.83224224357102\\
7.97276505860274	5.77671132680702\\
7.98865465115154	5.71853621187703\\
8.0045174958825	5.65832891808119\\
8.01962796416757	5.59457085052173\\
8.03473663191994	5.52882032257481\\
8.04882503160557	5.46137227626483\\
8.06253519965102	5.39166112706744\\
8.07522838732887	5.31895493162149\\
8.08657366480032	5.24474194330057\\
8.09658883928233	5.1676879521338\\
8.10581118519349	5.08908504141711\\
8.11529096690746	5.00817216378909\\
8.12520966184533	4.92494344212321\\
8.13471306372439	4.84017338350991\\
8.14281589450768	4.75414227902758\\
8.15064039635074	4.66723200445991\\
8.15706313531407	4.57943530073991\\
8.16354920614295	4.49160983969415\\
8.16926158082654	4.40393227273474\\
8.1744930553164	4.31542304179814\\
8.17847921776104	4.22719141330531\\
8.18164802464726	4.14009861206032\\
8.1830503215154	4.05354174349971\\
8.18453007705655	3.96702317370087\\
8.18487871539408	3.88155367990196\\
8.1839379809475	3.79664376687453\\
8.18217094186396	3.712622099953\\
8.17916731640544	3.62941058737184\\
8.17569389381948	3.54837968520431\\
8.17144442836054	3.46861959749976\\
8.16568413925905	3.38948427372301\\
8.15852296541455	3.3110086731817\\
8.15041028930217	3.23398424103764\\
8.1409820693201	3.15879997116761\\
8.13051877260217	3.08628283157761\\
8.11855471715271	3.01490058112001\\
8.10515249631196	2.94515781471642\\
8.09125184195626	2.87634347766468\\
8.07557641149851	2.80905996732272\\
8.05901030096639	2.74290733704193\\
8.04156006699496	2.6795774711473\\
8.02428528394902	2.61794114818865\\
8.0065809532266	2.55855360935992\\
7.98809409645775	2.50127066284259\\
7.96909138686323	2.44653099259768\\
7.94907099142181	2.39358683203726\\
7.9280283781611	2.34189270424254\\
7.90725233250464	2.2927858109104\\
7.88569380694937	2.24477317379716\\
7.86312676890134	2.19893357541186\\
7.84063949651005	2.15508831167047\\
7.81923471833377	2.11288317349388\\
7.79848455520224	2.07230062575276\\
7.77732746505362	2.03253868824919\\
7.75634675434418	1.99448384892862\\
7.73463169500801	1.9584549866464\\
7.71281632984263	1.92469251111738\\
7.69050393933314	1.89327469233256\\
7.6677838346924	1.86302565201948\\
7.64487697441893	1.83312804228392\\
7.62314832185006	1.80366785671299\\
7.60151368119774	1.77659703885802\\
7.57933900616676	1.75128805938269\\
7.5584340316474	1.72799635002207\\
7.53837112778426	1.70543318825948\\
7.51839681739248	1.68391892889235\\
};

\addplot [color=mycolor2]
  table[row sep=crcr]{%
0	4\\
0	4\\
0.005	3.998\\
0.0143	3.99428\\
0.027288	3.9890848\\
0.04342908	3.982628368\\
0.0622558328	3.97509766688\\
0.083359982048	3.9666560071808\\
0.10638503873568	3.95744598450573\\
0.131019867522989	3.9475920529908\\
0.156993050202603	3.93720277991896\\
0.184067947572025	3.92637282097119\\
0.212038373209323	3.91518465071627\\
0.240724803362255	3.9037100786551\\
0.269971056547358	3.89201157738106\\
0.299641384679822	3.88014344612807\\
0.329617924760647	3.86815283009574\\
0.359798466460796	3.85608061341568\\
0.390094496473403	3.84396220141064\\
0.420429485351324	3.83182820585947\\
0.450737386793389	3.81970504528264\\
0.480961323062862	3.80761547077486\\
0.511052433481022	3.79557902660759\\
0.540968865794514	3.78361245368219\\
0.570674892717155	3.77173004291314\\
0.600140138139037	3.75994394474439\\
0.629338899416422	3.74826444023343\\
0.658249553838694	3.73670017846452\\
0.686854038843016	3.72525838446279\\
0.715137396839055	3.71394504126438\\
0.743087376637963	3.70276504934481\\
0.770694084471345	3.69172236621146\\
0.797949678454778	3.68082012861809\\
0.824848101111588	3.67006075955536\\
0.851384845239535	3.65944606190419\\
0.877556748987346	3.64897730040506\\
0.903361816519984	3.63865527339201\\
0.928799061100078	3.62848037555997\\
0.95386836780592	3.61845265287763\\
0.978570373450743	3.6085718506197\\
1.00290636156968	3.59883745537213\\
1.02687817060506	3.58924873175797\\
1.05048811365235	3.57980475453906\\
1.07373890833181	3.57050443666727\\
1.09663361552884	3.56134655378846\\
1.11917558590163	3.55232976563935\\
1.14136841319116	3.54345263472353\\
1.16321589348836	3.53471364260466\\
1.18472198971757	3.52611120411297\\
1.20589080068771	3.51764367972492\\
1.22672653414259	3.50930938634296\\
1.24723348331242	3.50110660667503\\
1.27165711138197	3.50652523664729\\
1.29970341352179	3.52384289201839\\
1.33071522020606	3.55215288682062\\
1.36430886012344	3.58810133594319\\
1.40012570632963	3.63324390769794\\
1.43735371105169	3.68603763501153\\
1.47642019865556	3.74337500559185\\
1.51720923837119	3.80697341411115\\
1.55911292790389	3.875071199688\\
1.60160661076771	3.94797062630601\\
1.64464705208981	4.02326656744398\\
1.68808650166782	4.0998808485087\\
1.73206135152322	4.1786821940655\\
1.77619045171752	4.25821429603026\\
1.82042512605028	4.33954336662518\\
1.86440243162339	4.4208242962303\\
1.9091736930486	4.50347579138251\\
1.95418213425632	4.58641944797958\\
1.99935936453876	4.66862436488487\\
2.04497615046186	4.75106986771041\\
2.09004213652238	4.83384682800982\\
2.13462917227404	4.91509125423699\\
2.17967934267403	4.99486267337238\\
2.22430474081935	5.07250838362368\\
2.26833445712231	5.14846995572691\\
2.31160171895853	5.22409960554242\\
2.3546944605642	5.29719286979165\\
2.39680859769247	5.36948459177138\\
2.43876046563738	5.43947529204703\\
2.4792733446228	5.50826082605552\\
2.51944681528793	5.57391989629705\\
2.55848414657976	5.63866487421818\\
2.59664099132141	5.70157094248427\\
2.63397106832431	5.76183140869623\\
2.67040933246496	5.82053239428762\\
2.70597289596395	5.87775964777105\\
2.74139709136731	5.93246421235779\\
2.77568294985698	5.98422050635706\\
2.8098980669861	6.03363812071545\\
2.84334289925287	6.080402737152\\
2.8756816810542	6.12415720789208\\
2.90649218816335	6.16523500934214\\
2.93637368165039	6.2044661733777\\
2.96489761060255	6.24206028275598\\
2.99305555389507	6.27738843944006\\
3.02115669884683	6.30936483774289\\
3.04809740323397	6.33944437632106\\
3.07471985185402	6.36738200280889\\
3.10095246409971	6.39293005076096\\
3.1273418096405	6.41643327125908\\
3.15388550564981	6.4369368423977\\
3.17974308887681	6.4550133893699\\
3.20592122267103	6.47100353518461\\
3.23222127528927	6.48461122036004\\
3.2578449308712	6.49605851722816\\
3.28371676014932	6.50350524391737\\
3.35917289990285	6.51791032127123\\
3.43415619343869	6.50918171187\\
3.45977939364026	6.50672865621805\\
3.48546931782254	6.50238905549828\\
3.51220660580764	6.49518852538779\\
3.53874985874492	6.48597305309367\\
3.56524704422165	6.47416355232759\\
3.59258740287422	6.45913277689704\\
3.62040871586317	6.44117275648891\\
3.64836452055405	6.41992418348111\\
3.6774728510619	6.39684919264252\\
3.70709974870593	6.371145809434\\
3.73752926105971	6.34355389410559\\
3.76861863415953	6.31361998574005\\
3.80040512357378	6.28147405957772\\
3.83270529423655	6.24615071604524\\
3.86645142005726	6.20790273797443\\
3.90112964306247	6.16723298784042\\
3.93691158754152	6.12456576482561\\
3.97371183471535	6.07955867624205\\
4.01128788974866	6.03072026264817\\
4.04968867077864	5.97973423433149\\
4.08902069999421	5.92592042659793\\
4.12900617318419	5.86895433145633\\
4.16946232485759	5.81029093023745\\
4.21024529894405	5.74803612753885\\
4.25257433967188	5.68255117132717\\
4.29635352879025	5.61519357842899\\
4.34050639537405	5.54499793723311\\
4.38596128378257	5.47234643561073\\
4.43300121207307	5.39747369726725\\
4.48063495570264	5.32109827494887\\
4.5293780683059	5.24213966730871\\
4.57931578080916	5.16170178484966\\
4.62984930157847	5.08003061029581\\
4.68043332930353	4.99637165208877\\
4.73161747498052	4.91153170743834\\
4.78292136229295	4.82566352085548\\
4.834154703529	4.73876656416736\\
4.88609546177539	4.65026235162824\\
4.93769367465516	4.56006967664085\\
4.9896907096982	4.46782184625706\\
5.04250980430386	4.37364354607654\\
5.09476658370758	4.27983757140341\\
5.1460433099028	4.18660099354833\\
5.19729303878151	4.09304293597692\\
5.247696111161	3.99960026403938\\
5.29702856514877	3.90741437985519\\
5.34464958560551	3.81566798260063\\
5.39050545854659	3.72418052660028\\
5.43569433371098	3.63312378477266\\
5.48065865499497	3.544301960622\\
5.52401132370873	3.45770238899858\\
5.56625339260907	3.37178409116417\\
5.60685848386473	3.28810551987212\\
5.6454651056373	3.20695866922864\\
5.68241102957201	3.12793670581174\\
5.71854514369379	3.04970595189303\\
5.75395009737439	2.97386241362095\\
5.78708884337959	2.89977389857026\\
5.81936165042899	2.82790269321173\\
5.84983535485959	2.75722516833819\\
5.87901212820487	2.68834762955753\\
5.90566917346044	2.62073565783469\\
5.9313378922549	2.55449828614821\\
5.95456102732233	2.48983175346839\\
5.97590824081154	2.42661220173646\\
5.99576645377816	2.36469954576273\\
6.01363712664519	2.30385190499815\\
6.02987821847637	2.24477386269306\\
6.04408636879119	2.18759625729293\\
6.05763767420506	2.13173835045412\\
6.0695285508075	2.07819768344305\\
6.08011975968818	2.02676671054212\\
6.0899309577733	1.97711633452224\\
6.10000000134939	1.92889345343437\\
6.10880074469359	1.88272079452677\\
6.11707552055412	1.8378082126412\\
6.12451810223051	1.79583766509661\\
6.13075928234334	1.7556309490298\\
6.13718808061008	1.71777285232643\\
6.14357081460198	1.68153179799351\\
6.14929283897739	1.64720585300536\\
6.15501051201753	1.61493023358333\\
6.16052012751691	1.58502071547524\\
6.1646482269318	1.5565242874558\\
6.16873579258248	1.53067931123581\\
6.17301190429938	1.50707769120775\\
6.17653600275944	1.48392189064758\\
6.17963098849386	1.46169126743144\\
6.18140706056163	1.44147210941278\\
6.18302842302948	1.42151564990422\\
};

\addplot[only marks, mark=diamond*, mark options={}, mark size=0.6584pt, color=blue, fill=blue] table[row sep=crcr]{%
x	y\\
0	4\\
0	4\\
0.005	3.998\\
0.0143	3.99428\\
0.027288	3.9890848\\
0.04342908	3.982628368\\
0.0622558328	3.97509766688\\
0.083359982048	3.9666560071808\\
0.10638503873568	3.95744598450573\\
0.131019867522989	3.9475920529908\\
0.156993050202603	3.93720277991896\\
0.184067947572025	3.92637282097119\\
0.212038373209323	3.91518465071627\\
0.240724803362255	3.9037100786551\\
0.269971056547358	3.89201157738106\\
0.299641384679822	3.88014344612807\\
0.329617924760647	3.86815283009574\\
0.359798466460796	3.85608061341568\\
0.390094496473403	3.84396220141064\\
0.420429485351324	3.83182820585947\\
0.450737386793389	3.81970504528264\\
0.480961323062862	3.80761547077486\\
0.511052433481022	3.79557902660759\\
0.540968865794514	3.78361245368219\\
0.570674892717155	3.77173004291314\\
0.600140138139037	3.75994394474439\\
0.629338899416422	3.74826444023343\\
0.658249553838694	3.73670017846452\\
0.686854038843016	3.72525838446279\\
0.715137396839055	3.71394504126438\\
0.743087376637963	3.70276504934481\\
0.770694084471345	3.69172236621146\\
0.797949678454778	3.68082012861809\\
0.824848101111588	3.67006075955536\\
0.851384845239535	3.65944606190419\\
0.877556748987346	3.64897730040506\\
0.903361816519984	3.63865527339201\\
0.928799061100078	3.62848037555997\\
0.95386836780592	3.61845265287763\\
0.978570373450743	3.6085718506197\\
1.00290636156968	3.59883745537213\\
1.02687817060506	3.58924873175797\\
1.05048811365235	3.57980475453906\\
1.07373890833181	3.57050443666727\\
1.09663361552884	3.56134655378846\\
1.11917558590163	3.55232976563935\\
1.14136841319116	3.54345263472353\\
1.16321589348836	3.53471364260466\\
1.18472198971757	3.52611120411297\\
1.20589080068771	3.51764367972492\\
1.22672653414259	3.50930938634296\\
1.24723348331242	3.50110660667503\\
1.27165711138197	3.50652523664729\\
1.29970341352179	3.52384289201839\\
1.33071522020606	3.55215288682062\\
1.36430886012344	3.58810133594319\\
1.40012570632963	3.63324390769794\\
1.43735371105169	3.68603763501153\\
1.47642019865556	3.74337500559185\\
1.51720923837119	3.80697341411115\\
1.55911292790389	3.875071199688\\
1.60160661076771	3.94797062630601\\
1.64464705208981	4.02326656744398\\
1.68808650166782	4.0998808485087\\
1.73206135152322	4.1786821940655\\
1.77619045171752	4.25821429603026\\
1.82042512605028	4.33954336662518\\
1.86440243162339	4.4208242962303\\
1.9091736930486	4.50347579138251\\
1.95418213425632	4.58641944797958\\
1.99935936453876	4.66862436488487\\
2.04497615046186	4.75106986771041\\
2.09004213652238	4.83384682800982\\
2.13462917227404	4.91509125423699\\
2.17967934267403	4.99486267337238\\
2.22430474081935	5.07250838362368\\
2.26833445712231	5.14846995572691\\
2.31160171895853	5.22409960554242\\
2.3546944605642	5.29719286979165\\
2.39680859769247	5.36948459177138\\
2.43876046563738	5.43947529204703\\
2.4792733446228	5.50826082605552\\
2.51944681528793	5.57391989629705\\
2.55848414657976	5.63866487421818\\
2.59664099132141	5.70157094248427\\
2.63397106832431	5.76183140869623\\
2.67040933246496	5.82053239428762\\
2.70597289596395	5.87775964777105\\
2.74139709136731	5.93246421235779\\
2.77568294985698	5.98422050635706\\
2.8098980669861	6.03363812071545\\
2.84334289925287	6.080402737152\\
2.8756816810542	6.12415720789208\\
2.90649218816335	6.16523500934214\\
2.93637368165039	6.2044661733777\\
2.96489761060255	6.24206028275598\\
2.99305555389507	6.27738843944006\\
3.02115669884683	6.30936483774289\\
3.04809740323397	6.33944437632106\\
3.07471985185402	6.36738200280889\\
3.10095246409971	6.39293005076096\\
3.1273418096405	6.41643327125908\\
3.15388550564981	6.4369368423977\\
3.17974308887681	6.4550133893699\\
3.20592122267103	6.47100353518461\\
3.23222127528927	6.48461122036004\\
3.2578449308712	6.49605851722816\\
3.28371676014932	6.50350524391737\\
3.35917289990285	6.51791032127123\\
3.43415619343869	6.50918171187\\
3.45977939364026	6.50672865621805\\
3.48546931782254	6.50238905549828\\
3.51220660580764	6.49518852538779\\
3.53874985874492	6.48597305309367\\
3.56524704422165	6.47416355232759\\
3.59258740287422	6.45913277689704\\
3.62040871586317	6.44117275648891\\
3.64836452055405	6.41992418348111\\
3.6774728510619	6.39684919264252\\
3.70709974870593	6.371145809434\\
3.73752926105971	6.34355389410559\\
3.76861863415953	6.31361998574005\\
3.80040512357378	6.28147405957772\\
3.83270529423655	6.24615071604524\\
3.86645142005726	6.20790273797443\\
3.90112964306247	6.16723298784042\\
3.93691158754152	6.12456576482561\\
3.97371183471535	6.07955867624205\\
4.01128788974866	6.03072026264817\\
4.04968867077864	5.97973423433149\\
4.08902069999421	5.92592042659793\\
4.12900617318419	5.86895433145633\\
4.16946232485759	5.81029093023745\\
4.21024529894405	5.74803612753885\\
4.25257433967188	5.68255117132717\\
4.29635352879025	5.61519357842899\\
4.34050639537405	5.54499793723311\\
4.38596128378257	5.47234643561073\\
4.43300121207307	5.39747369726725\\
4.48063495570264	5.32109827494887\\
4.5293780683059	5.24213966730871\\
4.57931578080916	5.16170178484966\\
4.62984930157847	5.08003061029581\\
4.68043332930353	4.99637165208877\\
4.73161747498052	4.91153170743834\\
4.78292136229295	4.82566352085548\\
4.834154703529	4.73876656416736\\
4.88609546177539	4.65026235162824\\
4.93769367465516	4.56006967664085\\
4.9896907096982	4.46782184625706\\
5.04250980430386	4.37364354607654\\
5.09476658370758	4.27983757140341\\
5.1460433099028	4.18660099354833\\
5.19729303878151	4.09304293597692\\
5.247696111161	3.99960026403938\\
5.29702856514877	3.90741437985519\\
5.34464958560551	3.81566798260063\\
5.39050545854659	3.72418052660028\\
5.43569433371098	3.63312378477266\\
5.48065865499497	3.544301960622\\
5.52401132370873	3.45770238899858\\
5.56625339260907	3.37178409116417\\
5.60685848386473	3.28810551987212\\
5.6454651056373	3.20695866922864\\
5.68241102957201	3.12793670581174\\
5.71854514369379	3.04970595189303\\
5.75395009737439	2.97386241362095\\
5.78708884337959	2.89977389857026\\
5.81936165042899	2.82790269321173\\
5.84983535485959	2.75722516833819\\
5.87901212820487	2.68834762955753\\
5.90566917346044	2.62073565783469\\
5.9313378922549	2.55449828614821\\
5.95456102732233	2.48983175346839\\
5.97590824081154	2.42661220173646\\
5.99576645377816	2.36469954576273\\
6.01363712664519	2.30385190499815\\
6.02987821847637	2.24477386269306\\
6.04408636879119	2.18759625729293\\
6.05763767420506	2.13173835045412\\
6.0695285508075	2.07819768344305\\
6.08011975968818	2.02676671054212\\
6.0899309577733	1.97711633452224\\
6.10000000134939	1.92889345343437\\
6.10880074469359	1.88272079452677\\
6.11707552055412	1.8378082126412\\
6.12451810223051	1.79583766509661\\
6.13075928234334	1.7556309490298\\
6.13718808061008	1.71777285232643\\
6.14357081460198	1.68153179799351\\
6.14929283897739	1.64720585300536\\
6.15501051201753	1.61493023358333\\
6.16052012751691	1.58502071547524\\
6.1646482269318	1.5565242874558\\
6.16873579258248	1.53067931123581\\
6.17301190429938	1.50707769120775\\
6.17653600275944	1.48392189064758\\
6.17963098849386	1.46169126743144\\
6.18140706056163	1.44147210941278\\
6.18302842302948	1.42151564990422\\
};

\addplot[area legend, draw=black, fill=blue, fill opacity=0.3]
table[row sep=crcr] {%
x	y\\
4	7\\
2	7\\
2	5\\
4	5\\
}--cycle;

\addplot[area legend, draw=black, fill=green, fill opacity=0.3]
table[row sep=crcr] {%
x	y\\
8	7\\
6	7\\
6	5\\
8	5\\
}--cycle;

\addplot[area legend, draw=black, fill=mycolor3, fill opacity=0.3]
table[row sep=crcr] {%
x	y\\
8	3\\
6	3\\
6	1\\
8	1\\
}--cycle;

\addplot[area legend, draw=black, fill=black, fill opacity=0.6]
table[row sep=crcr] {%
x	y\\
6	3.5\\
8	3.5\\
8	4.5\\
6	4.5\\
}--cycle;
\node[right, align=left]
at (axis cs:-0.11368482,2.8336007503) {\textcolor{black}{\tiny$t=0$}};
\node[right, align=left]
at (axis cs:-0.11368482,4.3336007503) {\textcolor{blue}{\tiny$p_1$}};
\node[right, align=left]
at (axis cs:3.71368482,1.436007503) {\textcolor{black}{\tiny$t=0$}};
\node[right, align=left]
at (axis cs:3.991368482,2.636007503) {\textcolor{green}{\tiny$p_2$}};
\draw [->, line width=0.1mm] (0.1,3.6) -- (.9,3.3);
\draw [->, line width=0.1mm] (5,1.7) -- (4.2,2.01);
\end{axis}
\end{tikzpicture}%

%% file: ro_con.tex
%
%
\definecolor{mycolor1}{rgb}{1.00000,1.00000,0.00000}%
\definecolor{mycolor2}{rgb}{1.00000,1.00000,0.60000}%
\pgfplotsset{tick label style={font=\tiny}}
\begin{tikzpicture}

\begin{axis}[%
width=2.9cm,
height=2.6cm,
at={(0.0cm,0.0cm)},
scale only axis,
xmin=0,
xmax=20,
ymin=-0.15,
ymax=0.15,
axis background/.style={fill=white},
xlabel={\tiny time(s)},
xmajorgrids,
ymajorgrids,
legend style={at={(0.734,0.176)}, anchor=south west, legend cell align=left, align=left, draw=white!15!black}
]
\addplot [color=black,line width=.8pt]
  table[row sep=crcr]{%
0	0\\
0.1	0\\
0.2	0\\
0.3	0\\
0.4	0\\
0.5	0\\
0.6	0\\
0.7	0\\
0.8	0\\
0.9	0\\
1	0\\
1.1	0\\
1.2	0\\
1.3	0\\
1.4	0\\
1.5	0\\
1.6	0\\
1.7	0\\
1.8	0\\
1.9	0\\
2	0\\
2.1	0\\
2.2	0\\
2.3	0\\
2.4	0\\
2.5	0\\
2.6	0\\
2.7	0\\
2.8	0\\
2.9	0\\
3	0\\
3.1	0\\
3.2	0\\
3.3	0\\
3.4	0\\
3.5	0\\
3.6	0\\
3.7	0\\
3.8	0\\
3.9	0\\
4	0\\
4.1	0\\
4.2	0\\
4.3	0\\
4.4	0\\
4.5	0\\
4.6	0\\
4.7	0\\
4.8	0\\
4.9	0\\
5	0\\
};

\addplot [color=green,line width=1.1pt]
  table[row sep=crcr]{%
5.1	-0.118326793337695\\
5.2	-0.117315549101086\\
5.3	-0.115806573193134\\
5.4	-0.113863497838563\\
5.5	-0.111504889517493\\
5.6	-0.108789813204684\\
5.7	-0.105782364033314\\
5.8	-0.10248119749289\\
5.9	-0.0989472912521182\\
6	-0.095212850394202\\
6.1	-0.0913230094305938\\
6.2	-0.0873232153098553\\
6.3	-0.08319000682456\\
6.4	-0.0789833704319727\\
6.5	-0.074739012843187\\
6.6	-0.0704529829909965\\
6.7	-0.0661278897775577\\
6.8	-0.0617954859967396\\
6.9	-0.0574409795019623\\
7	-0.0531192200979927\\
7.1	-0.0488250337624786\\
7.2	-0.0445401471736402\\
7.3	-0.0403068865289772\\
7.4	-0.0360918006895403\\
7.5	-0.0319006527745556\\
7.6	-0.0277666954325206\\
7.7	-0.0237310583935893\\
7.8	-0.0197744030104809\\
7.9	-0.0158933921410253\\
8	-0.0127981554995248\\
8.1	-0.0110944223157921\\
8.2	-0.00942101891452607\\
8.3	-0.00779766273300242\\
8.4	-0.00619427031125313\\
8.5	-0.00461460803661227\\
8.6	-0.00305249744616887\\
8.7	-0.00155306836908553\\
8.8	-0.000108635038110072\\
8.9	0.0976898986848227\\
9	0.0979962487686803\\
9.1	0.0982641637324844\\
9.2	0.0984973335822539\\
9.3	0.0986988935037696\\
9.4	0.0988726487218223\\
9.5	0.0990226348881393\\
9.6	0.0991506521336973\\
9.7	0.09925952984155\\
9.8	0.0993516097293667\\
9.9	0.0994290948185232\\
10	0.0994927112751427\\
10.1	0.0995464219696458\\
10.2	0.0995892834417953\\
10.3	0.0996231541242598\\
10.4	0.0996511211929501\\
10.5	0.0996742276841325\\
10.6	0.099694467382442\\
10.7	0.0997138212282138\\
10.8	0.099732053489398\\
10.9	0.0997491065918585\\
11	0.0997654763726159\\
11.1	0.0997830226406868\\
11.2	0.0998004494509137\\
11.3	0.0998188012131065\\
11.4	0.0998371153384736\\
11.5	0.0998568590155027\\
11.6	0.09987666505947\\
11.7	0.0998970333317633\\
11.8	0.0999167933695984\\
11.9	0.0999352994529232\\
12	0.0999528477519327\\
12.1	0.0999683347191165\\
12.2	0.0999815645617845\\
12.3	0.0999908527932234\\
12.4	0.0999954719589278\\
12.5	0.099994325596434\\
12.6	0.0999858035083792\\
12.7	0.0999686932609589\\
12.8	0.0999418731446822\\
12.9	0.0999035813413676\\
13	0.0998518343294639\\
13.1	0.0997838754634373\\
13.2	0.0996984165485835\\
13.3	0.0995926993466396\\
13.4	0.0994643810095137\\
13.5	0.0993120336571198\\
13.6	0.0991299558094498\\
13.7	0.0989186234214408\\
13.8	0.0986772952962762\\
13.9	0.0984016058990715\\
14	0.098086583489394\\
14.1	0.0977354264118024\\
14.2	-0.00160665125411457\\
14.3	-0.00373232564892689\\
14.4	-0.00591847655983502\\
14.5	-0.00816410658964324\\
14.6	-0.0104498692893086\\
14.7	-0.012768138290685\\
14.8	-0.0151089646375334\\
14.9	-0.0174715857687139\\
15	-0.0174715857687139\\ 
};

\addplot [color=mycolor1,line width=.8pt]
  table[row sep=crcr]{%
15.1	-0.0278059888083133\\
15.2	-0.0254291646532359\\
15.3	-0.0230615201929124\\
15.4	-0.0207241459144512\\
15.5	-0.0184039399753335\\
15.6	-0.0160903073943639\\
15.7	-0.0138079738349956\\
15.8	-0.0115466977411081\\
15.9	-0.0093154322164292\\
16	-0.00711374332700702\\
16.1	-0.00497511567575425\\
16.2	-0.00287999340448394\\
16.3	-0.000814884445849939\\
16.4	0.0978042168141096\\
16.5	0.0981421413567567\\
16.6	0.0984436314416999\\
16.7	0.0987080170227963\\
16.8	0.0989428869530742\\
16.9	0.0991483752336133\\
17	0.0993272538022612\\
17.1	0.0994807537977964\\
17.2	0.0996108174552888\\
17.3	0.0997167158950205\\
17.4	0.0998021765040766\\
17.5	0.0998690695187745\\
17.6	0.0999199053253892\\
17.7	0.0999563587679233\\
17.8	0.0999809102547153\\
17.9	0.0999950978483488\\
18	0.099999910676879\\
18.1	0.0999969160867522\\
18.2	0.0999871896606548\\
18.3	0.0999719500846949\\
18.4	0.0999523472740602\\
18.5	0.0999292669531304\\
18.6	0.0999025144951711\\
18.7	0.0998734455104897\\
18.8	0.0998426099239536\\
18.9	0.0998111022548347\\
19	0.0997793931345452\\
19.1	0.0997464693613597\\
19.2	0.0997114730983573\\
19.3	0.0996754001202673\\
19.4	0.0996406395030622\\
19.5	0.0996061197443157\\
19.6	0.0995738763377647\\
19.7	0.0995417066432775\\
19.8	0.0995096630669363\\
19.9	0.0994782139977048\\
20	0.0994782139977048\\
};

\addplot [color=black, dashed,line width=1.3pt]
  table[row sep=crcr]{%
0	0\\
0.1	0\\
0.2	0\\
0.3	0\\
0.4	0\\
0.5	0\\
0.6	0\\
0.7	0\\
0.8	0\\
0.9	0\\
1	0\\
1.1	0\\
1.2	0\\
1.3	0\\
1.4	0\\
1.5	0\\
1.6	0\\
1.7	0\\
1.8	0\\
1.9	0\\
2	0\\
2.1	0\\
2.2	0\\
2.3	0\\
2.4	0\\
2.5	0\\
2.6	0\\
2.7	0\\
2.8	0\\
2.9	0\\
3	0\\
3.1	0\\
3.2	0\\
3.3	0\\
3.4	0\\
3.5	0\\
3.6	0\\
3.7	0\\
3.8	0\\
3.9	0\\
4	0\\
4.1	0\\
4.2	0\\
4.3	0\\
4.4	0\\
4.5	0\\
4.6	0\\
4.7	0\\
4.8	0\\
4.9	0\\
5	0\\ 
};

\addplot [color=mycolor2, dashed,line width=1.5pt]
  table[row sep=crcr]{%
15.1	-0.0346324768404264\\
15.2	-0.0302766193269695\\
15.3	-0.0259435908290925\\
15.4	-0.0216668064157726\\
15.5	-0.0174780621878222\\
15.6	-0.0133716529346919\\
15.7	-0.0105675755589611\\
15.8	-0.00840950353890634\\
15.9	-0.00631227076438903\\
16	-0.00424104698154064\\
16.1	-0.00223313723391204\\
16.2	-0.000296515022648436\\
16.3	0.0973111713111658\\
16.4	0.0977541903851371\\
16.5	0.0981430773659975\\
16.6	0.0984811540405726\\
16.7	0.0987733927380163\\
16.8	0.099024880933539\\
16.9	0.0992380560051769\\
17	0.0994164370336317\\
17.1	0.0995642753262493\\
17.2	0.0996833278844371\\
17.3	0.09977729405679\\
17.4	0.0998492073691879\\
17.5	0.099901750461995\\
17.6	0.0999367188426548\\
17.7	0.0999566585215996\\
17.8	0.0999636455577775\\
17.9	0.0999596967482344\\
18	0.0999468647613828\\
18.1	0.0999266039343329\\
18.2	0.0998993971943267\\
18.3	0.0998678838399312\\
18.4	0.0998322857055229\\
18.5	0.0997950488251445\\
18.6	0.0997569428711784\\
18.7	0.0997178611099261\\
18.8	0.0996780468259422\\
18.9	0.0996390935167117\\
19	0.0996009248458471\\
19.1	0.0995646989463301\\
19.2	0.0995310013027551\\
19.3	0.0995002477746927\\
19.4	0.0994718811130288\\
19.5	0.0994443486652445\\
19.6	0.099417944411121\\
19.7	0.0993957245071964\\
19.8	0.0993730423068406\\
19.9	0.0993491633130859\\
20	0.0993491633130859\\
};

\addplot [color=blue, dashed,line width=1.1pt]
  table[row sep=crcr]{%
5.1	-0.055507177660105\\
5.2	-0.0543026281430163\\
5.3	-0.0527194070006439\\
5.4	-0.0508614912629756\\
5.5	-0.0486942948645517\\
5.6	-0.0462814886369763\\
5.7	-0.0436993774242211\\
5.8	-0.0409044386643633\\
5.9	-0.0379607529768884\\
6	-0.0348731314967856\\
6.1	-0.0317088589134545\\
6.2	-0.0285007893984533\\
6.3	-0.0252240028362904\\
6.4	-0.0219266815830545\\
6.5	-0.0185833379524778\\
6.6	-0.0152515764185076\\
6.7	-0.0118711136742064\\
6.8	-0.00936810537104209\\
6.9	-0.00728404207442868\\
7	-0.00520146794003897\\
7.1	-0.00311872874055628\\
7.2	-0.00108346357930684\\
7.3	0.0967365143143397\\
7.4	0.0972172313530368\\
7.5	0.0976489862136487\\
7.6	0.0980384882441745\\
7.7	0.0983819390415435\\
7.8	0.0986826302948682\\
7.9	0.0989444566958335\\
8	0.099168016706934\\
8.1	0.0993572362288069\\
8.2	0.0995158376587666\\
8.3	0.0996457384797935\\
8.4	0.0997492914872922\\
8.5	0.0998306898889159\\
8.6	0.0998924166951849\\
8.7	0.0999367018644639\\
8.8	0.0999653083448933\\
8.9	0.0999813319481853\\
9	0.0999864136426485\\
9.1	0.0999825874279063\\
9.2	0.0999716855747985\\
9.3	0.0999549889488047\\
9.4	0.0999335176624803\\
9.5	0.0999086918908674\\
9.6	0.0998823923440899\\
9.7	0.0998545009264069\\
9.8	0.0998257329670693\\
9.9	0.0997969773010259\\
10	0.0997680553237865\\
10.1	0.0997405108550888\\
10.2	0.0997144681583118\\
10.3	0.0996890633514669\\
10.4	0.0996649082682357\\
10.5	0.0996424107273173\\
10.6	0.0996234985005122\\
10.7	0.0996073143008507\\
10.8	0.0995939244193893\\
10.9	0.0995826148735905\\
11	0.0995709349727891\\
11.1	0.0995584141059707\\
11.2	0.0995448841529574\\
11.3	0.0995302993315128\\
11.4	0.0995131636398814\\
11.5	0.0994925481675568\\
11.6	0.0994682090120738\\
11.7	0.0994356048138656\\
11.8	0.0993952773160545\\
11.9	0.0993446712982506\\
12	0.0992828751790935\\
12.1	0.0992080806577813\\
12.2	0.0991191551791508\\
12.3	0.0990070620715091\\
12.4	0.0988704765783199\\
12.5	0.0987063639882535\\
12.6	0.098512172926013\\
12.7	0.0982846943057072\\
12.8	0.0980216654982344\\
12.9	0.0977189553522193\\
13	-0.000333032730436877\\
13.1	-0.0018208405818289\\
13.2	-0.00334640218156157\\
13.3	-0.00493638097299638\\
13.4	-0.006586031595363\\
13.5	-0.00826715503414768\\
13.6	-0.0100002841133126\\
13.7	-0.0117906532313667\\
13.8	-0.0136125772603592\\
13.9	-0.0154772790504345\\
14	-0.0190146910146995\\
14.1	-0.0229929958186936\\
14.2	-0.0270321814298094\\
14.3	-0.0311253752336102\\
14.4	-0.0352560927381201\\
14.5	-0.0394185818280334\\
14.6	-0.0436445750710693\\
14.7	-0.0479092279069821\\
14.8	-0.052238438180014\\
14.9	-0.0566386944318957\\
15	-0.0389796466089122\\
};

\end{axis}
\end{tikzpicture}%

%% file: ro_con_obs.tex
%
%
\definecolor{mycolor1}{rgb}{0.00000,0.44700,0.74100}%
\definecolor{mycolor2}{rgb}{1.00000,1.00000,0.00000}%
\definecolor{mycolor3}{rgb}{1.00000,1.00000,0.50000}%
\pgfplotsset{tick label style={font=\tiny}}
\begin{tikzpicture}

\begin{axis}[%
width=2.9cm,
height=2.6cm,
at={(0.0cm,0.0cm)},
scale only axis,
xmin=0,
xmax=20,
ymin=-0.15,
ymax=0.15,
xlabel={\tiny time(s)},
xmajorgrids,
ymajorgrids,
axis background/.style={fill=white},
legend style={at={(0.751,0.193)}, anchor=south west, legend cell align=left, align=left, draw=white!15!black}
]

\addplot [color=black,line width=.8pt]
  table[row sep=crcr]{%
0	0\\
0.1	0\\
0.2	0\\
0.3	0\\
0.4	0\\
0.5	0\\
0.6	0\\
0.7	0\\
0.8	0\\
0.9	0\\
1	0\\
1.1	0\\
1.2	0\\
1.3	0\\
1.4	0\\
1.5	0\\
1.6	0\\
1.7	0\\
1.8	0\\
1.9	0\\
2	0\\
2.1	0\\
2.2	0\\
2.3	0\\
2.4	0\\
2.5	0\\
2.6	0\\
2.7	0\\
2.8	0\\
2.9	0\\
3	0\\
3.1	0\\
3.2	0\\
3.3	0\\
3.4	0\\
3.5	0\\
3.6	0\\
3.7	0\\
3.8	0\\
3.9	0\\
4	0\\
4.1	0\\
4.2	0\\
4.3	0\\
4.4	0\\
4.5	0\\
4.6	0\\
4.7	0\\
4.8	0\\
4.9	0\\
5	0\\
};

\addplot [color=green,line width=1.1pt]
  table[row sep=crcr]{%
5.1	-0.118328295988779\\
5.2	-0.117322584142529\\
5.3	-0.115815533964019\\
5.4	-0.113871737297131\\
5.5	-0.111512175162673\\
5.6	-0.108797271874951\\
5.7	-0.105784983951047\\
5.8	-0.102477983136928\\
5.9	-0.0989379515239409\\
6	-0.0951951783145936\\
6.1	-0.0912909398896086\\
6.2	-0.0872702844844459\\
6.3	-0.0831117835271966\\
6.4	-0.0788736219055097\\
6.5	-0.0745910406244048\\
6.6	-0.0702605736036337\\
6.7	-0.0658847740817282\\
6.8	-0.0614958552575307\\
6.9	-0.0570792151484133\\
7	-0.0526883090926736\\
7.1	-0.048317246972168\\
7.2	-0.0439482403450957\\
7.3	-0.0396229502449392\\
7.4	-0.0353080905062009\\
7.5	-0.0310088048601116\\
7.6	-0.0267584302790619\\
7.7	-0.0225984348856828\\
7.8	-0.0185085980328894\\
7.9	-0.0144843102423673\\
8	-0.0112971660321544\\
8.1	-0.00948745774242747\\
8.2	-0.00770086116062025\\
8.3	-0.00595675361940653\\
8.4	-0.00422268108752362\\
8.5	-0.00250139019696882\\
8.6	-0.000785300076405093\\
8.7	0.0974393377564786\\
8.8	0.0978395440289892\\
8.9	0.0981903345831865\\
9	0.0984953555931698\\
9.1	0.0987593980604164\\
9.2	0.0989859421913599\\
9.3	0.0991779542164668\\
9.4	0.0993388974494598\\
9.5	0.0994722995964124\\
9.6	0.0995798331341364\\
9.7	0.0996638588298271\\
9.8	0.0997271026355786\\
9.9	0.099771207671467\\
10	0.0997979345270221\\
10.1	0.099810302106478\\
10.2	0.0998089147468861\\
10.3	0.0997959882818047\\
10.4	0.0997751449412241\\
10.5	0.0997471900689739\\
10.6	0.0997117154781708\\
10.7	0.0996724423818316\\
10.8	0.0996291818412594\\
10.9	0.0995816045187969\\
11	0.0995312372774462\\
11.1	0.0994808676828682\\
11.2	0.099427133089361\\
11.3	0.0993743808232992\\
11.4	0.0993191905708091\\
11.5	0.0992650947706646\\
11.6	0.0992111651013006\\
11.7	0.0991591687739852\\
11.8	0.099110115274432\\
11.9	0.0990608263279713\\
12	0.0990150942691106\\
12.1	0.0989694028835268\\
12.2	0.0989236673787506\\
12.3	0.0988753733479719\\
12.4	0.0988261880537478\\
12.5	0.0987765646184697\\
12.6	0.0987220142314813\\
12.7	0.0986648736275006\\
12.8	0.0986039049130609\\
12.9	0.0985368789503698\\
13	0.0984583992779602\\
13.1	0.0983665227737656\\
13.2	0.0982656689864523\\
13.3	0.098151566914398\\
13.4	0.0980210023422281\\
13.5	0.0978708679146743\\
13.6	0.0976997504429391\\
13.7	0.0975048418213342\\
13.8	0.0972894746516078\\
13.9	0.0970478720555352\\
14	0.0967784958381401\\
14.1	0.0964851118223056\\
14.2	0.0961612096545454\\
14.3	-0.000145279854140723\\
14.4	-0.000382274396376792\\
14.5	-0.00250665571613968\\
14.6	-0.0048634922278548\\
14.7	-0.00721684060891081\\
14.8	-0.00958521001860417\\
14.9	-0.0119406960851218\\
15	-0.0119406960851218\\
};

\addplot [color=mycolor2,line width=.8pt]
  table[row sep=crcr]{%
15.1	-0.0368298465586953\\
15.2	-0.034747902646988\\
15.3	-0.032641765995253\\
15.4	-0.030543666135762\\
15.5	-0.0284148018429367\\
15.6	-0.0262888314859894\\
15.7	-0.024160810098958\\
15.8	-0.0220145439116188\\
15.9	-0.0198698262610108\\
16	-0.0177144478095455\\
16.1	-0.0156018396902427\\
16.2	-0.0135016008606972\\
16.3	-0.0113792105382903\\
16.4	-0.00923829117820152\\
16.5	-0.00710986347135425\\
16.6	-0.00499455122462212\\
16.7	-0.00292004031650163\\
16.8	-0.00083638266840943\\
16.9	-0.000128812618981478\\
17	0.0960140191431869\\
17.1	0.0963427587633885\\
17.2	0.0966496859148078\\
17.3	0.0969304087179039\\
17.4	0.097185904685007\\
17.5	0.0974187648650495\\
17.6	0.0976324913858999\\
17.7	0.0978266009250739\\
17.8	0.098006155537508\\
17.9	0.0981730460815511\\
18	0.0983213472577038\\
18.1	0.0984583004977304\\
18.2	0.0985842085816782\\
18.3	0.098696570973686\\
18.4	0.0987935012028067\\
18.5	0.0988776311323338\\
18.6	0.0989534454384322\\
18.7	0.0990195504125555\\
18.8	0.0990795780358737\\
18.9	0.0991326166921245\\
19	0.0991807844372088\\
19.1	0.099223973551656\\
19.2	0.0992613726196869\\
19.3	0.0992898960956483\\
19.4	0.0993145388229364\\
19.5	0.099336886853447\\
19.6	0.099353932741491\\
19.7	0.0993658130065633\\
19.8	0.0993748795830225\\
19.9	0.0993804411934329\\
20	0.0993804411934329\\
};

\addplot [color=black, dashed,line width=1.3pt]
  table[row sep=crcr]{%
0	0\\
0.1	0\\
0.2	0\\
0.3	0\\
0.4	0\\
0.5	0\\
0.6	0\\
0.7	0\\
0.8	0\\
0.9	0\\
1	0\\
1.1	0\\
1.2	0\\
1.3	0\\
1.4	0\\
1.5	0\\
1.6	0\\
1.7	0\\
1.8	0\\
1.9	0\\
2	0\\
2.1	0\\
2.2	0\\
2.3	0\\
2.4	0\\
2.5	0\\
2.6	0\\
2.7	0\\
2.8	0\\
2.9	0\\
3	0\\
3.1	0\\
3.2	0\\
3.3	0\\
3.4	0\\
3.5	0\\
3.6	0\\
3.7	0\\
3.8	0\\
3.9	0\\
4	0\\
4.1	0\\
4.2	0\\
4.3	0\\
4.4	0\\
4.5	0\\
4.6	0\\
4.7	0\\
4.8	0\\
4.9	0\\
5	0\\
};

\addplot [color=mycolor3, dashed,line width=1.3pt]
  table[row sep=crcr]{%
15.1	-0.0535139419382815\\
15.2	-0.0498937472764886\\
15.3	-0.0462976036680319\\
15.4	-0.0427596452132103\\
15.5	-0.0392754597699138\\
15.6	-0.0358418765458724\\
15.7	-0.0324357361205748\\
15.8	-0.029091082484219\\
15.9	-0.025842276475308\\
16	-0.0226382673064657\\
16.1	-0.0195311757423068\\
16.2	-0.0165373389314468\\
16.3	-0.0136381417472056\\
16.4	-0.0107790200457495\\
16.5	-0.00865124740597207\\
16.6	-0.00782277875541266\\
16.7	-0.00701595857875499\\
16.8	-0.00625411596757426\\
16.9	-0.00552469663353314\\
17	-0.00485827050174146\\
17.1	-0.00421655253148401\\
17.2	-0.00363597415440871\\
17.3	-0.00310229381679508\\
17.4	-0.00260583849225255\\
17.5	-0.00215907167020577\\
17.6	-0.00175304437406123\\
17.7	-0.00139784061583168\\
17.8	-0.00105905798013162\\
17.9	-0.000761786064723036\\
18	-0.000497005842364229\\
18.1	-0.000251725889899745\\
18.2	0.0977087274437858\\
18.3	0.0977290711455727\\
18.4	0.0977379487073629\\
18.5	0.0977363874273673\\
18.6	0.0977232133521007\\
18.7	0.0977063618408247\\
18.8	0.097684947286484\\
18.9	0.0976575782301667\\
19	0.097628273362464\\
19.1	0.0975976550227886\\
19.2	0.0975597746409491\\
19.3	0.0975236131927617\\
19.4	0.0974901645031647\\
19.5	0.0974519665050568\\
19.6	0.0974117273532036\\
19.7	0.0973687765799771\\
19.8	0.0973239881834704\\
19.9	0.0972807327343619\\
20	0.0972807327343619\\
};

\addplot [color=blue, dashed,line width=1.1pt]
  table[row sep=crcr]{%
5.1	-0.0555454412992685\\
5.2	-0.0544113423607504\\
5.3	-0.0529282973222394\\
5.4	-0.0511897450944083\\
5.5	-0.0491657596431687\\
5.6	-0.0469152163397626\\
5.7	-0.044505119882415\\
5.8	-0.0418954336736295\\
5.9	-0.0391453967928587\\
6	-0.0362605690527007\\
6.1	-0.0333021594880406\\
6.2	-0.0303008162187942\\
6.3	-0.0272314113303142\\
6.4	-0.0241398812731827\\
6.5	-0.0210007876468704\\
6.6	-0.0178693317643396\\
6.7	-0.0146837628468828\\
6.8	-0.0114849603988019\\
6.9	-0.00830040671621199\\
7	-0.00622325330723965\\
7.1	-0.00415382929975441\\
7.2	-0.00212271864407534\\
7.3	-0.000128433165690423\\
7.4	0.0966720504057463\\
7.5	0.0971314582258505\\
7.6	0.0975519402677378\\
7.7	0.097929028761883\\
7.8	0.0982682006239592\\
7.9	0.0985691198982541\\
8	0.0988335509869025\\
8.1	0.0990621441411701\\
8.2	0.0992599603996069\\
8.3	0.0994276474486868\\
8.4	0.0995664741283624\\
8.5	0.0996798381779436\\
8.6	0.0997695094672704\\
8.7	0.0998376139903832\\
8.8	0.0998850569253995\\
8.9	0.0999152858308689\\
9	0.0999295279211951\\
9.1	0.0999298385448335\\
9.2	0.0999180718803518\\
9.3	0.0998957730846457\\
9.4	0.0998640094172165\\
9.5	0.0998249751836235\\
9.6	0.0997814033596154\\
9.7	0.0997327807294091\\
9.8	0.0996804372145059\\
9.9	0.0996257808487833\\
10	0.0995688161614858\\
10.1	0.0995120454351976\\
10.2	0.0994557564859619\\
10.3	0.0993991211531495\\
10.4	0.0993433438473716\\
10.5	0.0992892595058363\\
10.6	0.0992404771670325\\
10.7	0.0991960349721639\\
10.8	0.0991565596744091\\
10.9	0.0991218114946717\\
11	0.0990882501794612\\
11.1	0.0990561002516051\\
11.2	0.0990253063553153\\
11.3	0.0989966195663272\\
11.4	0.0989689546122814\\
11.5	0.098941602963543\\
11.6	0.0989152105757194\\
11.7	0.0988834599242931\\
11.8	0.0988484207730185\\
11.9	0.0988069778579765\\
12	0.0987591634180887\\
12.1	0.0987034508830649\\
12.2	0.0986402572074241\\
12.3	0.0985643247487549\\
12.4	0.0984749513214414\\
12.5	0.0983694512219773\\
12.6	0.0982465800922299\\
12.7	0.0981056053339269\\
12.8	0.0979435215387552\\
12.9	0.0977576862249858\\
13	-0.000725154468814271\\
13.1	-0.00173655826141212\\
13.2	-0.00275613261432368\\
13.3	-0.00381435863325734\\
13.4	-0.00490883836194234\\
13.5	-0.00601266002725105\\
13.6	-0.00714903223816594\\
13.7	-0.00832503044611914\\
13.8	-0.00951587403753742\\
13.9	-0.0107344518532871\\
14	-0.0119828946665259\\
14.1	-0.013246232686405\\
14.2	-0.0146015420779481\\
14.3	-0.0180021443367591\\
14.4	-0.021431446184757\\
14.5	-0.0248847036334667\\
14.6	-0.0283958279036997\\
14.7	-0.0319406001009649\\
14.8	-0.0355467217372117\\
14.9	-0.0392216566074332\\
15	-0.0412216566074332\\
};

\end{axis}
\end{tikzpicture}%

%% file: roo.tex
%
%
\definecolor{mycolor1}{rgb}{1.00000,1.00000,0.00000}%
\pgfplotsset{tick label style={font=\tiny}}
\begin{tikzpicture}

\begin{axis}[%
width=4.5cm,
height=1.7cm,
at={(0cm,0cm)},
scale only axis,
xmin=0,
xmax=45,
xtick={ 0,  5, 15, 25, 35, 45},
ymin=-0.1,
ymax=0.2,
ytick={-0.1,0,0.1,0.2},
axis background/.style={fill=white},
xlabel={\tiny time(s)},
xmajorgrids,
ymajorgrids,
legend style={legend cell align=left, align=left, draw=white!15!black}
]
\addplot [color=black, line width=1.1pt]
  table[row sep=crcr]{%
0	0\\
0.1	0\\
0.2	0\\
0.3	0\\
0.4	0\\
0.5	0\\
0.6	0\\
0.7	0\\
0.8	0\\
0.9	0\\
1	0\\
1.1	0\\
1.2	0\\
1.3	0\\
1.4	0\\
1.5	0\\
1.6	0\\
1.7	0\\
1.8	0\\
1.9	0\\
2	0\\
2.1	0\\
2.2	0\\
2.3	0\\
2.4	0\\
2.5	0\\
2.6	0\\
2.7	0\\
2.8	0\\
2.9	0\\
3	0\\
3.1	0\\
3.2	0\\
3.3	0\\
3.4	0\\
3.5	0\\
3.6	0\\
3.7	0\\
3.8	0\\
3.9	0\\
4	0\\
4.1	0\\
4.2	0\\
4.3	0\\
4.4	0\\
4.5	0\\
4.6	0\\
4.7	0\\
4.8	0\\
4.9	0\\
5	0\\
};

\addplot [color=blue, line width=1.1pt]
  table[row sep=crcr]{%
5.1	-0.054999559382147\\
5.2	-0.0530033960684654\\
5.3	-0.0510163321181264\\
5.4	-0.0489869418261351\\
5.5	-0.0469576805559024\\
5.6	-0.044997109236679\\
5.7	-0.0429398731660451\\
5.8	-0.0409314186284915\\
5.9	-0.0388229953946301\\
6	-0.0368979313089556\\
6.1	-0.0349244704332506\\
6.2	-0.0330058311954817\\
6.3	-0.0310447712484384\\
6.4	-0.0290571992632217\\
6.5	-0.027023329619776\\
6.6	-0.0250951213206132\\
6.7	-0.0230496239209919\\
6.8	-0.0210204591364008\\
6.9	-0.0190976790043756\\
7	-0.017181730684995\\
7.1	-0.0153034956125027\\
7.2	-0.0133365410038791\\
7.3	-0.0113660989829164\\
7.4	-0.00950960071255348\\
7.5	-0.00767724665521471\\
7.6	-0.00582409519879725\\
7.7	-0.00397528490528097\\
7.8	-0.00198819603842648\\
7.9	-0.000122352443082885\\
8	0.0979850552717012\\
8.1	0.0983100511233994\\
8.2	0.0986027657078172\\
8.3	0.0988427329964334\\
8.4	0.0990545255873092\\
8.5	0.09923074078437\\
8.6	0.0993757080476949\\
8.7	0.0995040986934621\\
8.8	0.0995998932547062\\
8.9	0.0996874689445704\\
9	0.0997604617281025\\
9.1	0.0998153073649701\\
9.2	0.0998597544151099\\
9.3	0.0998984386060624\\
9.4	0.0999272983118948\\
9.5	0.0999481850987951\\
9.6	0.0999642580214797\\
9.7	0.0999734516130046\\
9.8	0.0999782170093635\\
9.9	0.0999799171972766\\
10	0.0999798174906026\\
10.1	0.0999791478965508\\
10.2	0.0999811151161463\\
10.3	0.0999822266804833\\
10.4	0.0999815977425724\\
10.5	0.0999803970472126\\
10.6	0.0999798198149366\\
10.7	0.0999810719146055\\
10.8	0.0999786739746513\\
10.9	0.0999801031513501\\
11	0.0999780449646885\\
11.1	0.0999750150921321\\
11.2	0.0999739478267372\\
11.3	0.099970690993862\\
11.4	0.0999698535329685\\
11.5	0.099967197129023\\
11.6	0.0999651443110361\\
11.7	0.099966332342238\\
11.8	0.099964564908118\\
11.9	0.099960914718066\\
12	0.0999619003140091\\
12.1	0.09996029073403\\
12.2	0.0999575174815814\\
12.3	0.0999569532902678\\
12.4	0.099951377785263\\
12.5	0.0999476936547308\\
12.6	0.0999448120190256\\
12.7	0.0999408888059969\\
12.8	0.0999413950602843\\
12.9	0.0999427193687632\\
13	0.0999395423589664\\
13.1	0.0999370080636981\\
13.2	0.0999355548382757\\
13.3	0.0999386714967372\\
13.4	0.0999391768621849\\
13.5	0.0999406798211049\\
13.6	0.099946737356523\\
13.7	0.0999493788207155\\
13.8	0.099954452408165\\
13.9	0.0999590527351462\\
14	0.0999615906486866\\
14.1	0.0999629394548309\\
14.2	0.099966060914221\\
14.3	0.099970209623597\\
14.4	0.099972664368867\\
14.5	0.0999699521895352\\
14.6	0.0999630115907257\\
14.7	0.0999561235631552\\
14.8	0.099941459130457\\
14.9	0.0999132533829858\\
15	0.0999132533829858\\
};

\addplot [color=green, line width=1.1pt]
  table[row sep=crcr]{%
15.1	0.0997629104880782\\
15.2	0.0998141134631696\\
15.3	0.0998528952680391\\
15.4	0.0998834728353406\\
15.5	0.0999063346572973\\
15.6	0.0999209285143579\\
15.7	0.0999372002489736\\
15.8	0.0999488163022719\\
15.9	0.0999589044442302\\
16	0.0999667186189119\\
16.1	0.0999711017852301\\
16.2	0.0999760853560328\\
16.3	0.0999812579753885\\
16.4	0.099981648826597\\
16.5	0.0999816204703583\\
16.6	0.0999831699943494\\
16.7	0.0999827242791287\\
16.8	0.0999830279927714\\
16.9	0.0999837207284802\\
17	0.099984517852072\\
17.1	0.0999838509906532\\
17.2	0.0999833802329415\\
17.3	0.0999833419545053\\
17.4	0.0999821002161638\\
17.5	0.0999826109661017\\
17.6	0.0999798093417259\\
17.7	0.0999771417597508\\
17.8	0.0999754234423442\\
17.9	0.0999740270010223\\
18	0.0999736288248039\\
18.1	0.0999745936761451\\
18.2	0.0999715823349359\\
18.3	0.099970644225694\\
18.4	0.0999697946494509\\
18.5	0.0999644580272696\\
18.6	0.099959121456445\\
18.7	0.0999596666714642\\
18.8	0.0999577774589135\\
18.9	0.0999529037281031\\
19	0.0999484261103549\\
19.1	0.0999481723801954\\
19.2	0.0999504429710678\\
19.3	0.0999484663670989\\
19.4	0.0999458051782078\\
19.5	0.0999496880726993\\
19.6	0.0999474382979095\\
19.7	0.0999467513183434\\
19.8	0.0999487136538952\\
19.9	0.0999464128861278\\
20	0.0999474903162436\\
20.1	0.0999536487941577\\
20.2	0.0999588822324635\\
20.3	0.0999635781913475\\
20.4	0.0999691892579835\\
20.5	0.0999703964776211\\
20.6	0.0999713862033678\\
20.7	0.0999721715430775\\
20.8	0.0999716442078373\\
20.9	0.0999722451574427\\
21	0.099971573757403\\
21.1	0.0999671071311512\\
21.2	0.0999605510045534\\
21.3	0.0999486905720026\\
21.4	0.0999303561344147\\
21.5	0.099910139148282\\
21.6	0.0998783768495077\\
21.7	0.0998350397000944\\
21.8	0.0997793512676939\\
21.9	0.0997091133621675\\
22	0.0996216341716447\\
22.1	0.0995272012377342\\
22.2	0.0994030559481527\\
22.3	0.099262913115576\\
22.4	0.0990987455154189\\
22.5	0.0989087596678533\\
22.6	0.0986723800691374\\
22.7	0.0984367275622011\\
22.8	0.0981637919358802\\
22.9	0.0978326480547917\\
23	0.0974671079745311\\
23.1	0.0971165849579614\\
23.2	-0.000113198142196813\\
23.3	-0.00193251353792454\\
23.4	-0.0034610136091856\\
23.5	-0.00496980479243954\\
23.6	-0.00670883739510795\\
23.7	-0.0083181090279818\\
23.8	-0.00986640966905552\\
23.9	-0.0115048884928554\\
24	-0.0129860395942824\\
24.1	-0.0144831676522147\\
24.2	-0.0158690281043825\\
24.3	-0.017255445725208\\
24.4	-0.0194080334642793\\
24.5	-0.0217527051970499\\
24.6	-0.0239077760419591\\
24.7	-0.0261355643074579\\
24.8	-0.02818690808053\\
24.9	-0.0303416583236658\\
25	-0.0220381552194947\\
};

\addplot [color=red, line width=1.1pt]
  table[row sep=crcr]{%
25.1	-0.0183855611316479\\
25.2	-0.0183855611316479\\
25.3	-0.0166820408258785\\
25.4	-0.0148183199567571\\
25.5	-0.013124294231271\\
25.6	-0.0111707637404207\\
25.7	-0.00948385722265235\\
25.8	-0.00763995034584795\\
25.9	-0.00584452245348199\\
26	-0.0040320625331344\\
26.1	-0.00230631574880078\\
26.2	-0.000609342885055653\\
26.3	0.0977525319988888\\
26.4	0.0980894588454198\\
26.5	0.0984188275948983\\
26.6	0.0986944854717926\\
26.7	0.0989171953661838\\
26.8	0.0991227603237195\\
26.9	0.0992998010258077\\
27	0.0994454448874402\\
27.1	0.0995565261776699\\
27.2	0.0996483989471122\\
27.3	0.0997328764767913\\
27.4	0.0998031371030401\\
27.5	0.0998615798602394\\
27.6	0.0999006567499365\\
27.7	0.0999354387223641\\
27.8	0.0999606744496422\\
27.9	0.0999772349185168\\
28	0.0999850208310951\\
28.1	0.0999893897428576\\
28.2	0.0999933435464275\\
28.3	0.0999938634270436\\
28.4	0.0999911411575951\\
28.5	0.099988444966421\\
28.6	0.0999854423024364\\
28.7	0.0999840469815061\\
28.8	0.0999813431007728\\
28.9	0.099980954050753\\
29	0.0999775607218536\\
29.1	0.09997658171091\\
29.2	0.0999750149523919\\
29.3	0.0999749952677396\\
29.4	0.0999759533528555\\
29.5	0.0999764005715915\\
29.6	0.0999799578004799\\
29.7	0.0999823981619232\\
29.8	0.0999850174549552\\
29.9	0.0999879003876967\\
30	0.0999887559044801\\
30.1	0.0999895526418566\\
30.2	0.0999920786472535\\
30.3	0.0999939782358841\\
30.4	0.0999958345895473\\
30.5	0.0999972403819935\\
30.6	0.0999972495640276\\
30.7	0.099997910032378\\
30.8	0.0999991981255273\\
30.9	0.0999996590862415\\
31	0.0999997636328733\\
31.1	0.0999998889627505\\
31.2	0.0999999369926323\\
31.3	0.0999997593653204\\
31.4	0.0999994379365596\\
31.5	0.0999991300705738\\
31.6	0.0999983387258501\\
31.7	0.0999973862093422\\
31.8	0.0999959701782311\\
31.9	0.0999943669000274\\
32	0.0999936341808965\\
32.1	0.0999922358421592\\
32.2	0.099990790427795\\
32.3	0.0999867918211834\\
32.4	0.099986805868878\\
32.5	0.099984091749679\\
32.6	0.0999825370544696\\
32.7	0.0999806316233744\\
32.8	0.0999767126458271\\
32.9	0.0999743617089082\\
33	0.0999711819240117\\
33.1	0.0999689367757806\\
33.2	0.099965894738836\\
33.3	0.0999642952684456\\
33.4	0.0999623714232982\\
33.5	0.0999602767895307\\
33.6	0.0999519941237401\\
33.7	0.0999487013463718\\
33.8	0.0999439419342922\\
33.9	0.0999379579183741\\
34	0.0999289063260811\\
34.1	0.0999182946431967\\
34.2	0.0999041874572315\\
34.3	0.0998845132066655\\
34.4	0.0998676967441219\\
34.5	0.0998448399433189\\
34.6	0.0998206246938518\\
34.7	0.0997862802552156\\
34.8	0.0997508031793508\\
34.9	0.0997038721201464\\
35	0.0997038721201464\\
};

\addplot [color=mycolor1, line width=1.1pt]
  table[row sep=crcr]{%
35.1	0.149293767800164\\
35.2	0.149387023802364\\
35.3	0.149464842479333\\
35.4	0.149546839293605\\
35.5	0.14960835229942\\
35.6	0.149669621273358\\
35.7	0.14972147658472\\
35.8	0.149774600530616\\
35.9	0.149819270081145\\
36	0.149858980332667\\
36.1	0.149888754872896\\
36.2	0.149910117897984\\
36.3	0.149930969638693\\
36.4	0.149944855889512\\
36.5	0.149960583107433\\
36.6	0.14996870075439\\
36.7	0.149974550261368\\
36.8	0.149981565165861\\
36.9	0.149985452438395\\
37	0.149989849354205\\
37.1	0.149993116432833\\
37.2	0.14999504588221\\
37.3	0.1499952865612\\
37.4	0.149996584032467\\
37.5	0.149997420902836\\
37.6	0.149997273334545\\
37.7	0.149996700416652\\
37.8	0.149996767432196\\
37.9	0.149995958495553\\
38	0.1499969094675\\
38.1	0.149996929523061\\
38.2	0.149996327009505\\
38.3	0.149996675820019\\
38.4	0.149997140891926\\
38.5	0.149997312954036\\
38.6	0.149998139959092\\
38.7	0.149997819979169\\
38.8	0.149997758489862\\
38.9	0.14999730996583\\
39	0.149997719156113\\
39.1	0.149997720684141\\
39.2	0.149996893072199\\
39.3	0.1499970297732\\
39.4	0.149997649112404\\
39.5	0.149998250684675\\
39.6	0.149997761729409\\
39.7	0.149997527910184\\
39.8	0.149997393081719\\
39.9	0.149996355181506\\
40	0.149996593293318\\
40.1	0.149997118029293\\
40.2	0.149996475345734\\
40.3	0.149997111990658\\
40.4	0.149997230725924\\
40.5	0.149997061238243\\
40.6	0.149996049957992\\
40.7	0.149996424514623\\
40.8	0.149995264317024\\
40.9	0.149993828291135\\
41	0.149993991003564\\
41.1	0.149994754696776\\
41.2	0.149993095439406\\
41.3	0.149994092987797\\
41.4	0.149993173819633\\
41.5	0.1499924821011\\
41.6	0.149993627811008\\
41.7	0.149992571671296\\
41.8	0.149992945049631\\
41.9	0.149991204353445\\
42	0.149991462422048\\
42.1	0.149990359814538\\
42.2	0.149989760054782\\
42.3	0.149991178038506\\
42.4	0.149990786675057\\
42.5	0.149992172222921\\
42.6	0.149992750968168\\
42.7	0.149993453375378\\
42.8	0.149992307290314\\
42.9	0.149993571002709\\
43	0.149993956015735\\
43.1	0.149993826904884\\
43.2	0.149994529895465\\
43.3	0.149994681589489\\
43.4	0.149995566137367\\
43.5	0.149994928770331\\
43.6	0.149996230641208\\
43.7	0.149997188325084\\
43.8	0.149997871637349\\
43.9	0.149998364268507\\
44	0.149999389379526\\
44.1	0.149999706383596\\
44.2	0.149999719280547\\
44.3	0.149999439942144\\
44.4	0.149999177897992\\
44.5	0.149998437311072\\
44.6	0.149997728371787\\
44.7	0.1499965348425\\
44.8	0.149994124605948\\
44.9	0.149993668612649\\
45	0.149993668612649\\
};
\node[right, align=left]
at (axis cs:5.011368482,.1336007503) {\textcolor{blue}{\tiny$p
_1 \in Blue$}};
\node[right, align=left]
at (axis cs:14.011368482,-.0513336007503) {\textcolor{green}{\tiny$p
_2 \in Green$}};
\node[right, align=left]
at (axis cs:25.1368482,.13436007503) {\textcolor{red}{\tiny$p
_3 \in Red$}};
\node[right, align=left]
at (axis cs:32.451368482,.078436007503) {\textcolor{yellow}{\tiny$p
_1 \in Yellow$}};

\end{axis}
\end{tikzpicture}%